\documentclass[a4paper,11pt]{article}
 
 \usepackage{amsmath, amssymb, latexsym, amscd, amsthm,amsfonts,amstext}
 \usepackage[mathscr]{eucal}
 \usepackage{graphicx}
 \usepackage{subfig}

 \newtheorem{theorem}{Theorem}[section]
 \newtheorem{lemma}[theorem]{Lemma}
 \newtheorem{corollary}[theorem]{Corollary}

\newtheorem{remark}{Remark}[section]

\title{ Recovering of dielectric constants of explosives via a globally
strictly convex cost functional\footnote{This research was supported by US Army Research Laboratory and US Army Research Office grants W911NF-11-1-0325 and W911NF-11-1-0399.}} 
\author{Michael V. Klibanov and Nguyen Trung Th\`anh \\
Department of Mathematics \& Statistics, \\ University of North Carolina at Charlotte, USA\\
Emails: \texttt{mklibanv@uncc.edu, tnguy152@uncc.edu}} 
\date{}

\begin{document}
\maketitle

\begin{abstract}
The inverse problem of estimating dielectric constants of explosives using
boundary measurements of one component of the scattered electric field is
addressed. It is formulated as a coefficient inverse problem for a
hyperbolic differential equation. After applying the Laplace transform, a
new cost functional is constructed and a variational problem is formulated.
The key feature of this functional is the presence of the Carleman Weight
Function for the Laplacian. The strict convexity of this functional on a
bounded set in a Hilbert space of an arbitrary size is proven. This allows
for establishing the global convergence of the gradient descent method. Some
results of numerical experiments are presented.

\end{abstract}

\textbf{Keywords}:  Coefficient inverse problem, Laplace transform, Carleman
weight function, strictly convex cost functional, global convergence,
Laguerre functions, numerical experiments.

\textbf{AMS classification codes:} 35R30, 35L05, 78A46

\def\bR{\mathbb{R}}
\def\bx{\mathbf{x}}

\section{Introduction}

\label{sec:1}

The detection and identification of explosive devices has always been of
particular interest to security and mine-countermeasures. One of the most
popular techniques used for the purpose of detection and identification is
the Ground Penetrating Radar (GPR). Exploiting the energy of backscattering
electromagnetic pulses measured on a boundary, the GPR allows for mapping
the internal structures containing explosives. Although this concept is not
new, combining the GPR with quantitative imaging may significantly enhance
the detection and especially identification of explosive devices. This idea
was recently proposed and developed in a series of publications. For
example, we mention \cite{BK,BNKF,NBKF,NBKF1}, where three-dimensional (3-d)
quantitative imaging of dielectric constants of targets mimicking explosives
was performed using experimental backscattering measurements; we also refer
to \cite{KBKSNF} for 1-d imaging using real data collected in the field by a
Forward Looking Radar.

In the above works, the imaging problem was formulated as a Coefficient
Inverse Problem (CIP) for a time dependent wave-like PDE. The main question
in performing quantitative imaging is as follows: \emph{How to find a good
approximation of the solution to the corresponding CIP without any advanced
knowledge of a small neighborhood of this solution?} We call a numerical
method providing such an approximation \emph{globally convergent}. As soon
as this approximation is found, a number of conventional locally convergent
methods may be used to refine that approximation, see, e.g., Chapters 4 and
5 of \cite{BK} for a two-stage numerical procedure.

It is well known that the objective functionals resulted from applying the
traditional least-squares method are usually nonconvex. This fact explains
existence of multiple local minima and ravines of those functionals. In
turn, the latter leads to the local convergence of gradient and Newton-like
methods. That is, the convergence of these methods is guaranteed only if an
initial approximation lies in a small neighborhood of the solution. However,
from our experience working with experimental data (see above citations), we
have observed that such a requirement is not normally satisfied in practical
situations.

In this paper we propose a new numerical method which provides the global
convergence. Currently there are two approaches to constructing globally
convergent algorithms. They were developed in a number of publications
summarized in books \cite{BK,KT}. In particular, the above cited works
treating experimental data use the approach of \cite{BK}. The common part of
both approaches is the temporal Laplace transform. Let $s>0$ be the
parameter in the Laplace transform. We assume that $s\geq \underline{s}$,
where $\underline{s}$ is a large enough positive constant. Then, applying
the Laplace transform to the time-dependent wave-like PDE, one obtains a
boundary value problem for a nonlinear integral differential equation.
However, the main difficulty then is that the integration over the parameter 
$s$ is required on the infinite interval $\left( s,\infty \right) $. In the
convexification method \cite{KT} that integral was truncated, and the
residual was set to zero. Unlike this, in \cite{BK} the residual, i.e., the
so-called \textquotedblleft tail function", was approximated in an iterative
process.

The open question is whether it is possible to avoid such a truncation. In
other words, whether it is possible to avoid a procedure of approximating
the tail function. This question is addressed in the current paper.

The \emph{main novelty} here is that the truncation of the infinite integral
is replaced by the truncation of a series with respect to some $s$-dependent
functions forming an orthonormal basis in $L_{2}\left( \underline{s},\infty
\right) .$ In this way, we obtain a coupled system of nonlinear elliptic
equations with respect to the $x-$dependent coefficients of the truncated
series, where $x\in \mathbb{R}^{3}$. In addition, we construct a least
squares objective functional with a Carleman Weight Function (CWF) in it.
The main result is formulated in Theorem \ref{th:1}. Given a bounded set of
an arbitrary size in a certain Hilbert space, one can choose a parameter of
the CWF such that this functional becomes strictly convex on this set. This
implies convergence of gradient methods in finding the unique minimizer of
the functional on that set, if it exists, starting from any point of that
set. Since restrictions on the diameter of that bounded set are not imposed,
we call the latter \emph{global convergence} and we call that functional 
\emph{globally} strictly convex. This idea also works in the case when the
Fourier transform is applied to the wave-like equation instead of the
Laplace transform.

To demonstrate the computational feasibility of the proposed method, we
perform a limited numerical study in the 1-d case. In particular, we show
numerically that if a locally convergent algorithm starts from an
approximation obtained by the proposed globally convergent algorithm, then
the accuracy of approximations can be significantly improved. On the other
hand, if that locally convergent algorithm starts from the background
medium, then it may fail to perform well. In section \ref{sec:2} we
formulate our inverse problem and theorems in 3-d. In sections \ref{sec:3}--%
\ref{sec:5} we prove these theorems. In section \ref{sec:1dprob} we briefly
summarize the special case of the 1-d problem. Section \ref{sec:6} describes
some details of the numerical implementation in the 1-d case. Finally, in
section \ref{sec:numexa} we demonstrate the numerical performance of the
proposed algorithm.

\section{The 3-d coefficient inverse problem}

\label{sec:2}

Below $x=\left( x_{1},x_{2},x_{3}\right) \in \mathbb{R}^{3}.$ Let $\Omega
\subset \mathbb{R}^{3}$ be a convex bounded domain with a piecewise-smooth
boundary $\partial \Omega $. We define $S_{\infty }:=\partial \Omega \times
\left( 0,\infty \right) .$ Let the function $c\left( x\right) $ satisfies
the following conditions%
\begin{equation}
c\in C^{1+\alpha }\left( \mathbb{R}^{3}\right) ,\ c_{0}\leq c\left( x\right)
\leq 1+d,\forall x\in \mathbb{R}^{3},\ c\left( x\right) =1,\ x\in \mathbb{R}%
^{3}\diagdown \Omega ,  \label{2.1}
\end{equation}%
where the numbers $c_{0}\in \left( 0,1\right) $ and $d>0$ are given. In this
paper, $C^{m+\alpha }$ denotes H\"{o}lder spaces, where $m\geq 0$ is an
integer and $\alpha \in \left( 0,1\right) $. Consider the following Cauchy
problem%
\begin{eqnarray}
&&c\left( x\right) u_{tt}=\Delta u\left( x,t\right) +\delta
(x_{3}-x_{3}^{0})f(t),\left( x,t\right) \in \mathbb{R}^{3}\times \left(
0,\infty \right) ,  \label{2.7} \\
&&u\left( x,0\right) =0,u_{t}\left( x,0\right) =0.  \label{2.8}
\end{eqnarray}%
The coefficient $c\left( x\right) $ represents the spatially distributed
dielectric constant. Here $c(x)$ is normalized so that its value in the
background medium, i.e., in $\mathbb{R}^{3}\setminus \Omega $, equals $1$.
The function $u\left( x,t\right) $ represents one of components of the
electric field generated by an incident plane wave propagating along the $%
x_{3}-$axis and excited at the plane $\left\{ x_{3}=x_{3}^{0}\right\} ,$
where $x_{3}^{0}\notin \overline{\Omega }$. The function $f\left( t\right)
\not\equiv 0$ is continuous and bounded which represents the time-dependent
waveform of the incident plane wave. Of course, the propagation of
electromagnetic waves should be modeled by the Maxwell's equations. However,
there are two reasons for us to use the scalar equation (\ref{2.7}). The
first reason is that most GPR systems provide only scalar data instead of
three components of the electric field. For example, in the experiments used
in \cite{BK,BNKF,KBKSNF,NBKF,NBKF1} only one component of the electric field
was generated by the source and the detector measured only that component of
the backscattering electric field. The second reason is that it was
demonstrated numerically in \cite{BM} that if the incident wave has only one
non-zero component of the electric field, then this component dominates the
other two. Moreover, equation (\ref{2.7}) approximates well the propagation
of that component even in 3-d \cite{BM}.

\textbf{CIP 1:} \emph{Suppose that conditions (\ref{2.1})--(\ref{2.8}) are
satisfied and that the plane }$\left\{ x_{3}=x_{3}^{0}\right\} \subset
\left( \mathbb{R}^{3}\setminus \overline{\Omega }\right) $\emph{. Determine
the coefficient }$c(x)$\emph{\ for }$x\in \Omega ,$\emph{\ assuming that the
following function }$p_{1}(x,t)$\emph{\ is known} 
\begin{equation}
u|_{S_{\infty }}=p_{1}(x,t).  \label{2.11}
\end{equation}

The function $p_{1}(x,t)$ in (\ref{2.11}) models a boundary measurement.
Having $p_{1}(x,t)$, one can uniquely solve the initial value problem (\ref%
{2.7})--(\ref{2.8}) outside of the domain $\Omega .$ Hence, the normal
derivative is also known: 
\begin{equation}
\partial _{n}u\mid _{S_{\infty }}=p_{2}\left( x,t\right) .  \label{2.12}
\end{equation}

We note that the knowledge of functions $p_{1}(x,t)$ and $p_{2}\left(
x,t\right) $ on an infinite rather than finite time interval is not a
serious restriction of our method, since the Laplace transform, which we
use, effectively cuts off values of these functions for large $t$. In
addition, if the incident plane wave is excited on a finite time interval,
the scattered wave will eventually vanish, as we observed in our experiments
in \cite{BK,BNKF,KBKSNF,NBKF,NBKF1}. In practice, incident waves are usually
excited for a short period of time.

\section{The coupled system of nonlinear elliptic equations}

\label{sec:3}

Consider the Laplace transform $\tilde{u}(x,s):=\left( \mathcal{L}u\right)
\left( x,s\right) =\int_{0}^{\infty }u\left( x,t\right) e^{-st}dt,\ s>0, $
where $s$ is referred to as the pseudofrequency. We also denote by $\tilde{f}%
(s)=\left( \mathcal{L}f\right) (s)$ the Laplace transform of $f(t)$.
Consider $s\geq \underline{s}(d)>0$, where the number $\underline{s}\left(
d\right) $ is large enough, so that the Laplace transforms of $u$ and its
derivatives $D^{\beta }u,\left\vert \beta \right\vert =1,2,$ converge
absolutely. The number $d$ is defined in (\ref{2.1}). We assume that $\tilde{%
f}(s)\neq 0$ for all $s\geq \underline{s}(d)$. Define $w(x,s):=\tilde{u}%
(x,s)/\tilde{f}(s)$. Then, this function satisfies the equation: 
\begin{equation}
\Delta w(x,s)-s^{2}c(x)w(x,s)=-\delta (x_{3}-x_{3}^{0}),\ x\in \mathbb{R}%
^{3},\ s\geq \underline{s}\left( d\right) .  \label{eq:w}
\end{equation}%
Define $w_{0}\left( x_{3},s\right) :=e^{-s|x_{3}-x_{3}^{0}|}/(2s).$ Note
that $w_{0}(x_{3},s)$ tends to zero as $\left\vert x_{3}\right\vert
\rightarrow \infty .$ The function $w_{0}\left( x_{3},s\right) $ is the
unique solution of equation (\ref{eq:w}) in the case $c(x)\equiv 1$ which
tends to zero as $\left\vert x_{3}\right\vert \rightarrow \infty .$ It is
shown in Theorem 3.1 of \cite{NBKF1} that in the case $f\left( t\right)
=\delta \left( t\right) $ 
\begin{equation}
\lim_{\left\vert x\right\vert \rightarrow \infty }\left[ w\left( x,s\right)
-w_{0}(x_{3},s)\right] =0  \label{1}
\end{equation}%
and that the function $w\left( x,s\right) $ can be represented in the form%
\begin{equation}
w\left( x,s\right) =w_{0}\left( x_{3},s\right) +\widehat{w}\left( x,s\right)
,\text{ where }\widehat{w}\left( x,s\right) \in C^{2+\alpha }\left( \mathbb{R%
}^{3}\right) ,\ \forall s\geq \underline{s}\left( d\right) .  \label{2}
\end{equation}%
Furthermore, the same theorem claims that $w(x,s)>0$ for all $s>\underline{s}%
\left( d\right) $. Thus, we assume these properties in our algorithm even if 
$f\left( t\right) \neq \delta \left( t\right) .$ Next, define $v:=(\ln
w)/s^{2}$. 
Substituting $w=e^{vs^{2}}$ into (\ref{eq:w}) and keeping in mind that $%
\overline{\Omega }\cap \{x_{3}=x_{3}^{0}\}=\varnothing $, we obtain 
\begin{equation}
\Delta v+s^{2}|\nabla v|^{2}=c(x),x\in \Omega .  \label{eq:c}
\end{equation}%
Hence, if the function $v$ is known, then the coefficient $c(x)$ can be
computed directly using (\ref{eq:c}). We define $q:=\partial v/\partial s$.
Thus, $v\left( x,s\right) =-\int\limits_{s}^{\infty }q\left( x,\tau \right)
d\tau . \label{3.9} $ Note that this converges absolutely together with its
derivatives with respect to $x$ up to the second order. The latter is true
if certain non-restrictive conditions are imposed on the function $c\left(
x\right) $ (see Lemma 6.5.2 in \cite{KT}), and we assume that these
conditions are in place. Hence, differentiating (\ref{eq:c}) with respect to 
$s$ leads to the following nonlinear integral differential equation as
mentioned in Introduction: 
\begin{equation}
\Delta q-2s^{2}\nabla q\int\limits_{s}^{\infty }\nabla q\left( x,\tau
\right) d\tau +2s\big(\int\limits_{s}^{\infty }\nabla q\left( x,\tau \right)
d\tau \big)^{2}=0,x\in \Omega ,s\geq \underline{s}\left( d\right) .
\label{3.10}
\end{equation}%
In addition, the following two boundary functions $\varphi \left( x,s\right) 
$, $\psi \left( x,s\right) $ can be derived from functions $p_{1}\left(
x,t\right) $, $p_{2}\left( x,t\right) $ in (\ref{2.11}) and (\ref{2.12})%
\begin{equation}
q\mid _{\partial \Omega }=\phi \left( x,s\right) ,\quad \partial _{n}q\mid
_{\partial \Omega }=\psi \left( x,s\right) ,\ s\geq \underline{s}\left(
d\right) .  \label{3.11}
\end{equation}

We have obtained the nonlinear boundary value problem (\ref{3.10})--(\ref%
{3.11}) for $q\left( x,s\right) .$ If this function is found, then the
coefficient $c\left( x\right) $ can be easily found via backwards
calculations. Therefore, the central focus should be on the solution of the
problem (\ref{3.10})--(\ref{3.11}).

We remark that functions $p_{1}\left( x,t\right) ,p_{2}\left( x,t\right) $
are results of measurements.\ Hence, they contain noise. Although one needs
to calculate the first derivative with respect to $s$ of functions $\left( 
\mathcal{L}p_{1}\right) \left( x,s\right) $, $\left( \mathcal{L}p_{2}\right)
\left( x,s\right) $ in order to find functions $\phi \left( x,s\right) $, $%
\psi \left( x,s\right) ,$ it was observed in \cite{BK} that this can be done
in a stable way, since the Laplace transform smooth out the that noise. In
addition, in our numerical computation we also remove high frequency noise
by truncating high order Fourier coefficients in the Fourier transformed
data.

There have been two globally convergent method proposed by our group so far 
\cite{BK,KT}. The common point of both methods is that the integral in (\ref%
{3.10}) is written as 
\begin{equation*}
\int\limits_{s}^{\infty }\nabla q(x,\tau )d\tau =\int\limits_{s}^{\bar{s}%
}\nabla q(x,\tau )d\tau +\nabla V(x,\bar{s}),\text{ }\nabla V(x,\bar{s}%
)=\int\limits_{\bar{s}}^{\infty }\nabla q(x,\tau )d\tau .
\label{eq:truncate}
\end{equation*}%
The function $V$ is called the \textquotedblleft tail
function\textquotedblright . In the method of \cite{KT} this tail function
was set to be zero, whereas in \cite{BK} it was approximated in an iterative
process.

The \emph{key novelty} of the method of this paper is that it does not
truncate the integral over $s$ in (\ref{3.10}) as in the above methods.
Instead, we represent the function $q(x,s)$ as a series with respect to an
orthonormal basis of $L_{2}(\underline{s},\infty )$. Using this
representation, the integral over the infinite interval $(s,\infty )$ in (%
\ref{3.10}) can be easily computed. Let $\left\{ f_{n}\left( s\right)
\right\} _{n=0}^{\infty }\subset L_{2}\left( \underline{s}\left( d\right)
,\infty \right) $ be an orthonormal basis in $L_{2}\left( \underline{s}%
\left( d\right) ,\infty \right) $ such that $\left\{ f_{n}\left( s\right)
\right\} _{n=0}^{\infty }\subset L_{1}\left( \underline{s}\left( d\right)
,\infty \right) .$ As an example, one can consider Laguerre functions \cite%
{AS:1964} 
\begin{equation*}
L_{n}\left( s\right) =e^{-s/2}\sum\limits_{k=0}^{n}\left( -1\right)
^{k}C_{n}^{k}\frac{s^{k}}{k!},\ s\in \left( 0,\infty \right) ,\ C_{n}^{k}=%
\frac{n!}{\left( n-k\right) !k!}.
\end{equation*}%
Next, we set $f_{n}\left( s\right) :=L_{n}\left( s-\underline{s}\left(
d\right) \right) ,s\in \left( \underline{s}\left( d\right) ,\infty \right) .$
It can be verified that $q\left( x,s\right) \in L_{2}\left( \underline{s}%
\left( d\right) ,\infty \right) ,\forall x\in \overline{\Omega }.$ Hence,
one can represent the function $q\left( x,s\right) $ as%
\begin{equation}
q\left( x,s\right) =\sum\limits_{n=0}^{\infty }q_{n}\left( x\right)
f_{n}\left( s\right) \approx \sum\limits_{n=0}^{N-1}q_{n}\left( x\right)
f_{n}\left( s\right) ,\text{ }s\geq \underline{s}\left( d\right) ,
\label{3.12}
\end{equation}%
where $N$ is a sufficiently large integer which should be chosen in
numerical experiments. Consider the vector of coefficients in the truncated
series (\ref{3.12}) $Q\left( x\right) =\left( q_{0},...,q_{N-1}\right)
\left( x\right) \in \mathbb{R}^{N}.$ Substituting the truncated series (\ref%
{3.12}) into (\ref{3.10}), we obtain 
\begin{equation}
\begin{split}
\sum\limits_{n=0}^{N-1}\Delta q_{n}(x)f_{n}(s)&
-2s^{2}\sum\limits_{m=0}^{N-1}\sum\limits_{n=0}^{N-1}\nabla q_{m}(x)\nabla
q_{n}(x)f_{m}(s)\int\limits_{s}^{\infty }f_{n}(\tau )d\tau \\
& +2s\sum\limits_{m=0}^{N-1}\sum\limits_{n=0}^{N-1}\nabla q_{m}(x)\nabla
q_{n}(x)\int\limits_{s}^{\infty }f_{m}(\tau )d\tau \int\limits_{s}^{\infty
}f_{n}(\tau )d\tau =0.
\end{split}
\label{eq:qn}
\end{equation}%
To be precise, one should have \textquotedblleft $\approx $" instead of
\textquotedblleft $=$" in (\ref{eq:qn}) due to the truncation (\ref{3.12}).
Multiplying both sides of (\ref{eq:qn}) by $f_{k}(s)$, integrating over $(%
\underline{s}\left( d\right) ,\infty )$ and keeping in mind the fact that $%
\left\{ f_{n}(s)\right\} _{n=0}^{\infty }$ is an orthonormal basis in $L_{2}(%
\underline{s},\infty )$, we obtain the following system of coupled nonlinear
elliptic equations: 
\begin{equation}
\Delta q_{k}(x)+\sum\limits_{m=0}^{N-1}\sum\limits_{n=0}^{N-1}F_{kmn}\nabla
q_{m}(x)\nabla q_{n}(x)=0,\ k=0,\dots ,N-1,\ x\in \Omega ,  \label{eq:qn2}
\end{equation}%
where the numbers $F_{kmn}$, $k,m,n\in \{0,\dots ,N-1\}$, are given by 
\begin{equation*}
F_{kmn}=\int\limits_{\underline{s}\left( d\right) }^{\infty }2sf_{k}(s)\big(%
\int\limits_{s}^{\infty }f_{m}(\tau )d\tau \int\limits_{s}^{\infty
}f_{n}(\tau )d\tau \big)ds-\int\limits_{\underline{s}\left( d\right)
}^{\infty }2s^{2}f_{k}(s)f_{m}(s)\big(\int\limits_{s}^{\infty }f_{n}(\tau
)d\tau \big)ds.
\end{equation*}%
The boundary conditions for $q_{n}$ are obtained by substituting again the
truncated series (\ref{3.12}) into (\ref{3.11}). For the convenience of the
following analysis, we rewrite system (\ref{eq:qn2}) together with the
boundary conditions as the following boundary value problem with
over-determined boundary conditions. Note that we have both Dirichlet and
Neumann boundary conditions 
\begin{eqnarray}
&&\Delta Q+F\left( \nabla Q\right) =0,  \label{3.13} \\
&&Q\mid _{\partial \Omega }=\Phi \left( x\right) ,\partial _{n}Q\mid
_{\partial \Omega }=\Psi \left( x\right) ,  \label{3.14}
\end{eqnarray}%
where the boundary vector functions $\Phi \left( x\right) ,\Psi \left(
x\right) \in \mathbb{R}^{N}$ are computed from the functions $\varphi \left(
x,s\right) ,\psi \left( x,s\right) $ and $F:\mathbb{R}^{3N}\rightarrow 
\mathbb{R}^{N}$, $F=\left( F_{0},...,F_{N-1}\right) \in C^{\infty }\left( 
\mathbb{R}^{3N}\right) $ with 
\begin{equation*}
F_{k}(\nabla Q)=\sum\limits_{m=0}^{N-1}\sum\limits_{n=0}^{N-1}F_{kmn}\nabla
q_{m}(x)\nabla q_{n}(x),\ k=0,\dots ,N-1.
\end{equation*}

If we can find an approximate solution of the problem (\ref{3.13})--(\ref%
{3.14}), then we can find an approximation for the function $q$ via the
truncated series (\ref{3.12}). Therefore, we focus below on the method of
approximating the vector function $Q\left( x\right) .$

Let $\widetilde{F}\left( x\right) =\big( \widetilde{F}_{0},...,\widetilde{F}%
_{N-1}\big) \left( x\right) ,x\in \Omega, $ be a vector function and $H$ be
a Hilbert space. Below any statement that $\widetilde{F}\in H$ means that
every component of the vector $\widetilde{F}$ belongs to $H$. The norm $%
\left\Vert \widetilde{F}\right\Vert _{H}$ means%
\begin{equation*}
\left\Vert \widetilde{F}\right\Vert _{H}=\big( \sum\limits_{n=0}^{N-1}\left%
\Vert \widetilde{F}_{n}\right\Vert _{H}^{2}\big) ^{1/2}.
\end{equation*}

\section{Globally Convex Cost Functional}

\label{sec:4}

Our ultimate goal is to apply this method to the inversion of experimental
data of \cite{BNKF,NBKF,NBKF1}. Thus, just as in these references, below $%
\Omega $ is chosen to be a rectangular parallelepiped. Without loss of
generality, it is convenient to assume that 
\begin{equation*}
\Omega =\left\{ x=\left( x_{1},x_{2},x_{3}\right) :\left( x_{1},x_{2}\right)
\in \left( -A,A\right) ,x_{3}\in \left( 0,1/2\right) \right\} ,
\end{equation*}%
where $A>0$ is a number. Thus, $\partial \Omega =\Gamma _{1}\cup \Gamma
_{2}\cup \Gamma _{3},$ where 
\begin{equation*}
\Gamma _{1}=\left\{ x\in \partial \Omega |x_{3}=0\right\} ,\ \Gamma
_{2}=\left\{ x\in \partial \Omega |x_{3}=1/2\right\} ,\ \Gamma _{3}=\Omega
\diagdown \left( \Gamma _{1}\cup \Gamma _{2}\right) .
\end{equation*}%
As in \cite{BNKF,NBKF,NBKF1}, $\Gamma _{1}$ is considered as the
backscattering side, where the data are measured. Although measurements were
not performed on $\Gamma _{2}\cup \Gamma _{3},$ it was demonstrated in these
references that assigning 
\begin{equation}
w(x,s)\mid _{\Gamma _{2}\cup \Gamma _{3}}:=w_{0}(x,s)\mid _{\Gamma _{2}\cup
\Gamma _{3}}  \label{4.0}
\end{equation}%
does not affect the accuracy of the reconstruction via the technique of \cite%
{BK}. This is probably because of the condition (\ref{1}). Thus, we now
relax conditions (\ref{3.14}), assuming that the normal derivative is given
only on $\Gamma _{1},$ the Dirichlet condition is given on $\Gamma _{1}\cup
\Gamma _{3}$ and no boundary condition is given on $\Gamma _{2}$, 
\begin{equation}
Q\mid _{\Gamma _{1}\cup \Gamma _{3}}=\Phi \left( x\right) ,\partial
_{n}Q\mid _{\Gamma _{1}}=\Psi \left( x\right) .  \label{4.1}
\end{equation}

Let us introduce a CWF for the Laplace operator which is suitable for this
domain $\Omega $ and for boundary conditions (\ref{4.1}). Let $a,\xi \in
\left( 0,1/2\right) $ be two arbitrary numbers. Let $\lambda ,\nu >1$ be two
large parameters which we will choose later. Then the CWF has the form 
\begin{equation}
\varphi _{\lambda ,\nu }\left( x_{3}\right) =e^{\lambda \left( x_{3}+\xi
\right) ^{-\nu }}e^{-\lambda \left( a+\xi \right) ^{-\nu }}.  \label{4.2}
\end{equation}%
Hence,%
\begin{equation}
\lim_{\lambda \rightarrow \infty }\varphi _{\lambda ,\nu }\left( 1/2\right)
=0.  \label{4.21}
\end{equation}%
Lemma \ref{le:1} establishes a Carleman estimate for the operator $\Delta $
in the domain $\Omega $ with the weight function (\ref{4.2}). The proof of
this lemma is almost identical to the proof of Lemma 6.5.1 of \cite{BK} and
is therefore omitted.

\begin{lemma}\label{le:1}
There exist sufficiently large
numbers $\lambda _{0}=\lambda _{0}\left( \Omega \right) >1,\nu _{0}=\nu
_{0}\left( \Omega ,a,\xi \right) >1$ depending only on the listed
parameters such that for an arbitrary function $u\in H^{2}\left( \Omega
\right) $ satisfying $u\mid _{\Gamma _{3}}=0$ the following
Carleman estimate holds for all $\lambda \geq \lambda _{0}$ and with
a constant $C=C\left( \Omega \right) >0$ depending only on the
domain $\Omega $ 
\begin{equation}
\begin{split}
& \int\limits_{\Omega }\left( \Delta u\right) ^{2}\varphi _{\lambda ,\nu
_{0}}^{2}dx+C\left( \left\Vert u\mid _{\Gamma _{1}}\right\Vert _{H^{1}\left(
\Gamma _{1}\right) }^{2}+\left\Vert \partial _{n}u\mid _{\Gamma
_{1}}\right\Vert _{L_{2}\left( \Gamma _{1}\right) }^{2}\right) e^{2\lambda
\xi ^{-\nu _{0}}} \\
& \geq C\int\limits_{\Omega }\left( \lambda \left\vert \nabla u\right\vert
^{2}+\lambda ^{3}u^{2}\right) \varphi _{\lambda ,\nu _{0}}^{2}dx-C\varphi
_{\lambda ,\nu _{0}}^{2}\left( 1/2\right) \int\limits_{\Gamma _{2}}\left(
\left\vert \nabla u\right\vert ^{2}+u^{2}\right) dx_{2}dx_{3}.
\end{split}
\label{4.20}
\end{equation}
\end{lemma}


Below $C=C\left( \Omega \right) >0$ denotes different constants depending
only on the domain $\Omega .$ Let $R>0$ be an arbitrary number. Define the
set $G$ of vector functions $Q$ as 
\begin{equation}
G=G\left( R,\Phi ,\Psi \right) =\left\{ 
\begin{array}{c}
Q=\left( q_{0},...,q_{N-1}\right) ^{T}\in H^{3}\left( \Omega \right)
:\left\Vert Q\right\Vert _{H^{3}\left( \Omega \right) }<R, \\ 
Q\mid _{\Gamma _{1}\cup \Gamma _{3}}=\Phi \left( x\right) ,\partial
_{n}Q\mid _{\Gamma _{1}}=\Psi \left( x\right) .%
\end{array}%
\right.  \label{4.3}
\end{equation}%
Then $G$ is an open set in $H^{3}\left( \Omega \right) .$ Also, Embedding
Theorem implies that 
\begin{equation}
G\subset C^{1}\left( \overline{\Omega }\right) ,\left\Vert Q\right\Vert
_{C^{1}\left( \overline{\Omega }\right) }<CR,\forall Q\in G.  \label{4.4}
\end{equation}%
Let $\nu _{0}=\left( \Omega ,a,\xi \right) >1$ be the number in Lemma \ref%
{le:1}. Denote $\Omega _{a}=\Omega \cap \left\{ x_{3}<a\right\} .$ We seek
the solution $Q$ of the problem (\ref{3.13}), (\ref{4.1}) on the set $G$ via
minimizing the following Tikhonov-like cost functional with the CWF $\varphi
_{\lambda ,\nu _{0}}^{2}$ and with the regularization parameter $\alpha \in
\left( 0,1\right) $ 
\begin{equation}
J_{\lambda ,\alpha }\left( Q\right) =\frac{1}{2}\int\limits_{\Omega }\left[
\Delta Q+F\left( \nabla Q\right) \right] ^{2}\varphi _{\lambda ,\nu
_{0}}^{2}dx+\frac{\alpha }{2}\left\Vert Q\right\Vert _{H^{3}\left( \Omega
\right) }^{2},Q\in G\left( R,\Phi ,\Psi \right) .  \label{4.5}
\end{equation}

\begin{theorem}\label{th:1}
There exists a sufficiently large number 
$\lambda _{1}=\lambda _{1}\left( \Omega ,G,F\right) >1$ depending
only on $\Omega ,G,F$ such that if $\lambda \geq \lambda _{1}$ 
 and $\alpha \in \left[ \varphi _{\lambda ,\nu _{0}}^{2}\left(
1/2\right) ,1\right) $, then the functional $J_{\lambda ,\alpha
}\left( Q\right) $ is strictly convex on the set $G$, i.e.,
there exists a constant $C_{1}=C_{1}\left( \Omega ,G,F\right) >0$
depending only on $\Omega ,G,F$ such that for all $Q_{1},Q_{2}\in G$
\begin{equation}
\begin{split}
& J_{\lambda ,\alpha }\left( Q_{2}\right) -J_{\lambda ,\alpha }\left(
Q_{1}\right) -J_{\lambda ,\alpha }^{\prime }\left( Q_{1}\right) \left(
Q_{2}-Q_{1}\right) \\
& \geq C_{1}\left\Vert Q_{2}-Q_{1}\right\Vert _{H^{1}\left( \Omega
_{a}\right) }^{2}+\frac{\alpha }{2}\left\Vert Q_{2}-Q_{1}\right\Vert
_{H^{3}\left( \Omega \right) }^{2},
\end{split}
\label{4.51}
\end{equation}%
where $J_{\lambda ,\alpha }^{\prime }\left( Q_{1}\right) $ is
the Fr\'{e}chet derivative of the functional $J_{\lambda ,\alpha }$
at the point $Q_{1}.$

\end{theorem}
%

\textbf{Proof}. The existence of the Fr\'{e}chet derivative of the
functional $J_{\lambda ,\alpha }$ is shown in the proof. Everywhere below $%
C_{1}=C_{1}\left( \Omega ,G,F\right) >0$ denotes different positive
constants depending only on the listed parameters. Denote $h=Q_{2}-Q_{1}.$
Then by (\ref{4.3})%
\begin{equation}
h\mid _{\Gamma _{1}\cup \Gamma _{3}}=0,h_{x_{3}}\mid _{\Gamma _{1}}=0.
\label{4.6}
\end{equation}%
Denote $H_{0}^{3}\left( \Omega \right) $ the subspace of the space $%
H^{3}\left( \Omega \right) $ consisting of vector functions satisfying
conditions (\ref{4.6}). Let $F^{\prime }\left( \nabla Q\right) $ be the $%
N\times N$ matrix, 
\begin{equation*}
F^{\prime }\left( \nabla Q\right) \left( x\right) =\left( \frac{\partial
F_{i}}{\partial q_{jx_{k}}}\left( \nabla Q\left( x\right) \right) \right)
_{\left( i,j,k\right) =\left( 1,0,1\right) }^{\left( N,N-1,3\right)
},q_{jx_{k}}\left( x\right) =\frac{\partial q_{j}\left( x\right) }{\partial
x_{k}}.
\end{equation*}%
Hence, (\ref{4.4}) implies that 
\begin{equation}
\left\vert F^{\prime }\left( \nabla Q\right) \left( x\right) \right\vert
\leq C_{1},\forall Q\in G\left( R,\Phi ,\Psi \right) ,\forall x\in \overline{%
\Omega }.  \label{4.61}
\end{equation}%
Next, by Taylor's formula 
\begin{equation*}
F\left( \nabla Q_{2}\right) :=F\left( \nabla Q_{1}+\nabla h\right) =F\left(
\nabla Q_{1}\right) +F^{\prime }\left( \nabla Q_{1}\right) \nabla h+P\left(
\nabla Q_{1},\nabla h\right) ,\forall x\in \overline{\Omega }.
\end{equation*}%
where 
\begin{equation}
\left\vert P\left( \nabla Q_{1},\nabla h\right) \right\vert \left( x\right)
\leq C_{1}\left\vert \nabla h\left( x\right) \right\vert ^{2},\forall x\in 
\overline{\Omega }.  \label{4.610}
\end{equation}%
Hence, for all $x\in \overline{\Omega }$ 
\begin{equation}
\begin{split}
& \left[ \left( \Delta Q_{1}+\Delta h\right) +F\left( \nabla Q_{1}+\nabla
h\right) \right] ^{2} \\
& =\left[ \Delta Q_{1}+F\left( \nabla Q_{1}\right) +\Delta h+F^{\prime
}\left( \nabla Q_{1}\right) \nabla h+P\left( \nabla Q_{1},\nabla h\right) %
\right] ^{2} \\
& =\left( \Delta Q_{1}+F\left( \nabla Q_{1}\right) \right) ^{2}+2\left(
\Delta Q_{1}+F\left( \nabla Q_{1}\right) \right) \left[ \Delta h+F^{\prime
}\left( \nabla Q_{1}\right) \nabla h\right] +\left( \Delta h\right) ^{2} \\
& +2P\left( \nabla Q_{1},\nabla h\right) \Delta h+P^{2}\left( \nabla
Q_{1},\nabla h\right) \\
& +2P\left( \nabla Q_{1},\nabla h\right) \left[ \Delta Q_{1}+F\left( \nabla
Q_{1}\right) +\Delta h+F^{\prime }\left( \nabla Q_{1}\right) \nabla h\right]
.
\end{split}
\label{0}
\end{equation}%
By (\ref{4.610}) and the Cauchy-Schwarz inequality $2P\left( \nabla
Q_{1},\nabla h\right) \Delta h\geq -\left( \Delta h\right)
^{2}/2-C_{1}\left\vert \nabla h\left( x\right) \right\vert ^{2}.$ Hence,
using (\ref{4.61})--(\ref{0}), we obtain%
\begin{equation}
\begin{split}
& \left[ \Delta Q_{1}+\Delta h+F\left( \nabla Q_{1}+\nabla h\right) \right]
^{2}-\left[ \Delta Q_{1}+F\left( \nabla Q_{1}\right) \right] ^{2} \\
& -2\left[ \Delta Q_{1}+F\left( \nabla Q_{1}\right) \right] \left[ \Delta
h+F^{\prime }\left( \nabla Q_{1}\right) \nabla h\right] \geq \frac{1}{2}%
\left( \Delta h\right) ^{2}-C_{1}\left( \nabla h\right) ^{2}.
\end{split}
\label{4.6100}
\end{equation}%
On the other hand,%
\begin{equation*}
\frac{\alpha }{2}\left\Vert Q_{1}+h\right\Vert _{H^{3}\left( \Omega \right)
}^{2}=\frac{\alpha }{2}\left\Vert Q_{1}\right\Vert _{H^{3}\left( \Omega
\right) }^{2}+\frac{\alpha }{2}\left\Vert h\right\Vert _{H^{3}\left( \Omega
\right) }^{2}+\alpha \left( Q_{1},h\right) _{H^{3}\left( \Omega \right) },
\end{equation*}%
where $\left( ,\right) _{H^{3}\left( \Omega \right) }$ is the scalar product
in $H^{3}\left( \Omega \right) .$ It follows from (\ref{0}) that 
\begin{equation}
J_{\lambda ,\alpha }^{\prime }\left( Q_{1}\right) h=\int\limits_{\Omega } 
\left[ \Delta Q_{1}+F\left( \nabla Q_{1}\right) \right] \left[ \Delta
h+F^{\prime }\left( \nabla Q_{1}\right) \nabla h\right] +\alpha \left(
Q_{1},h\right) _{H^{3}\left( \Omega \right) }.  \label{4.62}
\end{equation}

The right hand side of (\ref{4.62}) is a bounded linear functional acting on
the function $h\in H_{0}^{3}\left( \Omega \right) .$ Hence, Riesz theorem
and (\ref{4.62}) imply that there exists an element $M\left( Q_{1}\right)
\in H_{0}^{3}\left( \Omega \right) $ such that $J_{\lambda ,\alpha }^{\prime
}\left( Q_{1}\right) h=\left( M\left( Q_{1}\right) ,h\right) _{H^{3}\left(
\Omega \right) },\forall h\in H_{0}^{3}\left( \Omega \right) .$ Thus, the Fr%
\'{e}chet derivative $J_{\lambda ,\alpha }^{\prime }\left( Q\right) $ of the
functional $J_{\lambda ,\alpha }\left( Q\right) $ exists and $J_{\lambda
,\alpha }^{\prime }\left( Q\right) =M\left( Q\right) ,\forall Q\in G\left(
R,\Phi ,\Psi \right) .$ By (\ref{0})--(\ref{4.62}) 
\begin{equation}
\begin{split}
& J_{\lambda ,\alpha }\left( Q_{1}+h\right) -J_{\lambda ,\alpha }\left(
Q_{1}\right) -J_{\lambda ,\alpha }^{\prime }\left( Q_{1}\right) h \\
& \geq \int\limits_{\Omega }\left[ \frac{1}{4}\left( \Delta h\right)
^{2}-C_{1}\left( \nabla h\right) ^{2}\right] \varphi _{\lambda ,\nu
_{0}}^{2}dx+\alpha \left\Vert h\right\Vert _{H^{3}\left( \Omega \right)
}^{2}.
\end{split}
\label{4.7}
\end{equation}%
For $x\in \Omega _{a},$ $\varphi _{\lambda ,\nu _{0}}^{2}\left( x\right)
\geq \varphi _{\lambda ,\nu _{0}}^{2}\left( a\right) =1.$ Hence, by Lemma %
\ref{le:1} for sufficiently large $\lambda \geq \lambda _{1}=\lambda
_{1}\left( \Omega ,G,F\right) >1$

\begin{equation}
\begin{split}
& \int\limits_{\Omega }\left[ \frac{1}{4}\left( \Delta h\right)
^{2}-C_{1}\left( \nabla h\right) ^{2}\right] \varphi _{\lambda ,\nu
_{0}}^{2}dx\geq C_{1}\int\limits_{\Omega }\left[ \lambda \left\vert \nabla
\left( h\right) \right\vert ^{2}+\lambda ^{3}h^{2}\right] \varphi _{\lambda
,\nu _{0}}^{2}dx \\
& -C_{1}\varphi _{\lambda ,\nu _{0}}^{2}\left( 1/2\right)
\int\limits_{\Gamma _{2}}\left( \left\vert \nabla h\right\vert
^{2}+h^{2}\right) dx_{2}dx_{3} \\
& \geq C_{1}\left\Vert Q_{2}-Q_{1}\right\Vert _{H^{1}\left( \Omega
_{a}\right) }^{2}-C_{1}\varphi _{\lambda ,\nu _{0}}^{2}\left( 1/2\right)
\left\Vert Q_{2}-Q_{1}\right\Vert _{H^{3}\left( \Omega \right) }^{2}.
\end{split}
\label{4.8}
\end{equation}%
By (\ref{4.21}) the lower boundary of $\alpha \geq \varphi _{\lambda ,\nu
_{0}}^{2}\left( 1/2\right) $ tends to zero as $\lambda \rightarrow \infty $
and also 
\begin{equation*}
\frac{\alpha }{2}\left\Vert Q_{2}-Q_{1}\right\Vert _{H^{3}\left( \Omega
\right) }^{2}\geq \frac{1}{2}\varphi _{\lambda ,\nu _{0}}^{2}\left(
1/2\right) \left\Vert Q_{2}-Q_{1}\right\Vert _{H^{3}\left( \Omega \right)
}^{2}.
\end{equation*}%
Hence, (\ref{4.7}) and (\ref{4.8}) imply (\ref{4.51}). $\hfill \square $

\begin{corollary}[Uniqueness]\label{co:1}
There exists at most one
vector function $Q\in G$ satisfying conditions (\ref{3.13}), (\ref%
{4.1}).
\end{corollary}


\textbf{Proof.} Let $\widetilde{J}_{\lambda ,\alpha }\left( Q\right)
=J_{\lambda ,\alpha }\left( Q\right) -\alpha \left\Vert Q\right\Vert
_{H^{3}\left( \Omega \right) }/2.$ Suppose that there exist two vector
functions $Q_{1},Q_{2}\in G$\emph{\ }satisfying conditions (\ref{3.13}), (%
\ref{4.1}). Then $\widetilde{J}_{\lambda ,\alpha }\left( Q_{1}\right) =%
\widetilde{J}_{\lambda ,\alpha }\left( Q_{2}\right) =0.$ On the other hand, $%
\widetilde{J}_{\lambda ,\alpha }\left( Q\right) \geq 0,\forall Q\in G.$
Hence, $Q_{1}$ and $Q_{2}$ are points of minimum of the functional $%
\widetilde{J}_{\lambda ,\alpha }\left( Q\right) .$ Hence, $\widetilde{J}%
_{\lambda ,\alpha }^{\prime }\left( Q_{1}\right) =\widetilde{J}_{\lambda
,\alpha }^{\prime }\left( Q_{2}\right) =0.$ Hence, repeating the proof of
Theorem \ref{th:1}, we obtain the following analog of (\ref{4.8}) 
\begin{equation}
\left\Vert Q_{2}-Q_{1}\right\Vert _{H^{1}\left( \Omega _{a}\right) }^{2}\leq
\varphi _{\lambda ,\nu _{0}}^{2}\left( 1/2\right) \left\Vert
Q_{2}-Q_{1}\right\Vert _{H^{3}\left( \Omega \right) }^{2}.  \label{4.9}
\end{equation}%
Setting in (\ref{4.9}) $\lambda \rightarrow \infty $ and using (\ref{4.21}),
we obtain $Q_{1}=Q_{2}$ in $\Omega _{a}.$ Since $a\in \left( 0,1/2\right) $
is an arbitrary number, then $Q_{1}=Q_{2}$ in $\Omega .$ $\hfill \square $

\section{Global convergence of the gradient descent method}

\label{sec:5} It is well-known that the gradient descent method is globally
convergent for functionals which are strictly convex on the entire space.
However, the functional (\ref{4.5}) is strictly convex only on the bounded
set $G\left( R,\Phi ,\Psi \right) $. Therefore we need to prove the global
convergence of this method on this set. Suppose that a minimizer $Q_{\min }$
of (\ref{4.5}) exists on $G\left( R,\Phi ,\Psi \right) $. In the
regularization theory $Q_{\min }$ is called \emph{regularized solution} of
the problem (\ref{3.13}), (\ref{4.1}) \cite{BK}. Theorem \ref{th:1}
guarantees that such a minimizer is unique. First, we estimate in this
section the distance between regularized and exact solutions, depending on
the level of error in the data. Next, we establish that Theorem \ref{th:1}
implies that the gradient descent method of the minimization of the
functional (\ref{4.5}) converges to $Q_{\min }$ if starting at any point of
this set, i.e., it converges globally. In addition, we estimate the distance
between points of the minimizing sequence of the gradient descent method and
the exact solution of the problem. In principle, global convergence of other
gradient methods for the functional (\ref{4.5}) can also be proved. However,
we are not doing this for brevity.

\subsection{The distance between regularized and exact solutions}

Following one of concepts of Tikhonov for ill-posed problems (see, e.g.,
section 1.4 in \cite{BK}), we assume that there exist noiseless boundary
data $\Phi ^{\ast }\left( x\right) $ and $\Psi ^{\ast }\left( x\right) $
which correspond to the exact solution $Q^{\ast }$ of the problem (\ref{3.13}%
), (\ref{4.1}). Also, we assume that functions $\Phi \left( x\right) $ and $%
\Psi \left( x\right) $ at the part $\Gamma _{1}$ of the boundary contain an
error of the level $\delta ,$%
\begin{equation}
\left\Vert \Phi -\Phi ^{\ast }\right\Vert _{H^{1}\left( \Gamma _{1}\right)
}\leq \delta ,\left\Vert \Psi -\Psi ^{\ast }\right\Vert _{L_{2}\left( \Gamma
_{1}\right) }\leq \delta .  \label{5.7}
\end{equation}%
On the other hand, we do not assume any error in the function $\Phi $ at $%
\Gamma _{2}\cup \Gamma _{3},$ see a heuristic condition (\ref{4.0}), which
was justified numerically in \cite{BNKF,NBKF,NBKF1}. Theorem \ref{th:2}
estimates the distance between $Q_{\min }$ and $Q^{\ast }$ in the norm of $%
H^{1}\left( \Omega _{a}\right) ,$ which might be sufficient for
computations. Note that while in Theorem \ref{th:1} we have compared
functions $Q_{1}$ and $Q_{2}$ satisfying the same boundary conditions,
functions $Q_{\min }$ and $Q^{\ast }$ in Theorem \ref{th:2} satisfy
different boundary conditions, because of the error in the data.

\begin{theorem}\label{th:2}
Assume that conditions of Theorem \ref%
{th:1} hold and $\lambda \geq \lambda _{1}$. In addition, assume
that conditions (\ref{5.7}) are satisfied and also $\Phi \mid _{\Gamma
_{3}}=\Phi ^{\ast }\mid _{\Gamma _{3}}$. Suppose that there exists an
exact solution $Q^{\ast }$ of the problem (\ref{3.13}), (\ref{4.1})
and $Q^{\ast }\in G\left( R,\Phi ^{\ast },\Psi ^{\ast }\right) .$ In
addition, assume that there exists a minimizer $Q_{\min }\in G\left( R,\Phi
,\Psi \right) $ of the functional $J_{\lambda ,\alpha }.$ Let
the number $\delta _{0}=\delta _{0}\left( \Omega ,G,F,a,\xi \right) \in
\left( 0,1\right) $ be so small that %
\begin{equation}
\delta _{0}^{-\xi ^{-\nu _{0}}/2}>\lambda _{1}.  \label{5.9}
\end{equation}%
Let $\delta \in \left( 0,\delta _{0}\right) .$ Choose the
regularization parameter $\alpha $ in (\ref{4.5}) as $\alpha
=\alpha \left( \delta \right) =\delta ^{2\gamma },$ where
\begin{equation*}
2\gamma =\frac{\xi ^{\nu _{0}}}{2\left( a+\xi \right) ^{\nu _{0}}}\left[ 1-%
\frac{\xi ^{\nu _{0}}}{\left( a+\xi \right) ^{\nu _{0}}}\right] \in \left( 0,%
\frac{1}{2}\right) .
\end{equation*}%
 Then $\alpha \in \left( \varphi _{\lambda ,\nu _{0}}^{2}\left(
1/2\right) ,1\right) $ (as in Theorem \ref{th:1}) and 
\begin{equation}
\left\Vert Q^{\ast }-Q_{\min }\right\Vert _{H^{1}\left( \Omega _{a}\right)
}\leq C_{1}\delta ^{\gamma },\text{ }\forall \delta \in \left( 0,\delta
_{0}\right) .  \label{5.10}
\end{equation}

\end{theorem}

\textbf{Proof}. Denote $\overline{h}=Q^{\ast }-Q_{\min }.$ In the proof of
Theorem \ref{th:1} the function $h=Q_{2}-Q_{1}$ satisfies zero boundary
conditions (\ref{4.6}). Now, however, the only zero condition for the
function $\overline{h}$ is $\overline{h}\mid _{\Gamma _{3}}=0.$ Still, it is
obvious that one can slightly modify the proof of Theorem \ref{th:1} for the
case of non-zero Dirichlet and Neumann boundary conditions for $\overline{h}$
at $\Gamma _{1}.$ To do so, we take into account the second term on the left
hand side of (\ref{4.20}). Thus, 
\begin{eqnarray}
&&\left( \left\Vert \overline{h}\mid _{\Gamma _{1}}\right\Vert _{H^{1}\left(
\Gamma _{1}\right) }^{2}+\left\Vert \partial _{n}\overline{h}\mid _{\Gamma
_{1}}\right\Vert _{L_{2}\left( \Gamma _{1}\right) }^{2}\right) e^{2\lambda
\xi ^{-\nu _{0}}}+J_{\lambda ,\alpha }\left( Q^{\ast }\right) -J_{\lambda
,\alpha }\left( Q_{\min }\right)   \notag \\
-J_{\lambda ,\alpha }^{\prime }\left( Q_{\min }\right) \overline{h} &\geq
&C_{1}\left\Vert Q_{2}-Q_{1}\right\Vert _{H^{1}\left( \Omega _{a}\right)
}^{2}+\frac{\alpha }{2}\left\Vert Q_{2}-Q_{1}\right\Vert _{H^{3}\left(
\Omega \right) }^{2},\forall \lambda \geq \lambda _{1}.  \label{5.12}
\end{eqnarray}%
By (\ref{5.7}) 
\begin{equation}
\left( \left\Vert \overline{h}\mid _{\Gamma _{1}}\right\Vert _{H^{1}\left(
\Gamma _{1}\right) }^{2}+\left\Vert \partial _{n}\overline{h}\mid _{\Gamma
_{1}}\right\Vert _{L_{2}\left( \Gamma _{1}\right) }^{2}\right) e^{2\lambda
\xi ^{-\nu _{0}}}\leq \delta ^{2}e^{2\lambda \xi ^{-\nu _{0}}}.  \label{5.13}
\end{equation}%
Since $\delta \in \left( 0,\delta _{0}\right) ,$ then it follows from (\ref%
{5.9}) that one can choose $\lambda =\lambda \left( \delta \right) >\lambda
_{1}$ such that $\delta ^{2}e^{2\lambda \xi ^{-\nu _{0}}}=\delta .$ Thus, $%
\lambda =\lambda \left( \delta \right) =\ln \delta ^{-\xi ^{\nu _{0}}/2}$.
It can be easily verified that the above choice $\alpha =\alpha \left(
\delta \right) =\delta ^{2\gamma }$ guarantees that $\alpha \in \left(
\varphi _{\lambda ,\nu _{0}}^{2}\left( 1/2\right) ,1\right) .$ Hence, (\ref%
{5.12}) and (\ref{5.13}) imply that for such $\lambda $ 
\begin{equation}
\left\Vert Q^{\ast }-Q_{\min }\right\Vert _{H^{1}\left( \Omega _{a}\right)
}^{2}\leq C_{1}\delta +J_{\lambda ,\alpha }\left( Q^{\ast }\right)
-J_{\lambda ,\alpha }\left( Q_{\min }\right) -J_{\lambda ,\alpha }^{\prime
}\left( Q_{\min }\right) \overline{h}.  \label{5.14}
\end{equation}

Next, since $Q^{\ast }$ is the exact solution of the problem (\ref{3.13}), (%
\ref{4.1}), then $\Delta Q^{\ast }+F\left( \nabla Q^{\ast }\right) =0$ in $%
\Omega .$ Hence, (\ref{4.5}) implies that 
\begin{equation}
J_{\lambda ,\alpha }\left( Q^{\ast }\right) =\frac{\alpha }{2}\left\Vert
Q^{\ast }\right\Vert _{H^{3}\left( \Omega \right) }^{2}\leq \alpha
R^{2}=\delta ^{2\gamma }R^{2}.  \label{5.15}
\end{equation}%
Finally, since $J_{\lambda ,\alpha }^{\prime }\left( Q_{\min }\right) 
\overline{h}=0,$ $-J_{\lambda ,\alpha }\left( Q_{\min }\right) \leq 0$ and $%
\delta <\delta ^{2\gamma },$ then (\ref{5.14}) and (\ref{5.15}) imply that $%
\left\Vert Q^{\ast }-Q_{\min }\right\Vert _{H^{1}\left( \Omega _{a}\right)
}\leq C_{1}\delta ^{\gamma },$ which establishes (\ref{5.10}). $\hfill
\square $

\subsection{Global convergence of the gradient descent method}

We now formulate the gradient descent method with the constant step size $%
\beta $ for the problem of the minimization of the functional $J_{\lambda
,\alpha }.$ For brevity we do not indicate the dependence of functions $%
Q_{n} $ on parameters $\lambda ,\alpha ,\beta $. Let $Q_{1}\in G\left(
R/4,\Phi ,\Psi \right) $ be an arbitrary point of the set $G\left( R/4,\Phi
,\Psi \right) $. Consider the sequence $\left\{ Q_{n}\right\} _{n=1}^{\infty
}$ of the gradient descent method, 
\begin{equation}
Q_{n+1}=Q_{n}-\beta J_{\lambda ,\alpha }^{\prime }\left( Q_{n}\right)
,n=1,2,...  \label{5.1}
\end{equation}

\begin{theorem}\label{th:3}
Choose parameters $\lambda _{1},\nu
_{0},\alpha $ as in Theorem \ref{th:1} and let $\lambda \geq \lambda
_{1}.$ Assume that the functional $J_{\lambda ,\alpha }$
achieves its minimal value on the set $G\left( R,\Phi ,\Psi \right) $
 at a point $Q_{\min }\in G\left( R/4,\Phi ,\Psi \right) $
Consider the sequence (\ref{5.1}), in which the starting point $Q_{1}$
\ is an arbitrary point of the set $G\left( R/4,\Phi ,\Psi \right) $. 
Then there exists a sufficiently small number $\beta =\beta \left( \lambda
,\alpha ,G\left( R,\Phi ,\Psi \right) \right) \in \left( 0,1\right) $ %
and a number $\theta =\theta \left( \beta \right) \in \left( 0,1\right) ,$%
 both dependent only on the listed parameters, such that the sequence
(\ref{5.1}) converges to the point $Q_{\min },$%
\begin{equation}
\left\Vert Q_{n+1}-Q_{\min }\right\Vert _{H^{3}\left( \Omega \right) }\leq
\theta ^{n}\left\Vert Q_{1}-Q_{\min }\right\Vert _{H^{3}\left( \Omega
\right) },n=1,2,...  \label{5.2}
\end{equation}%
In addition, assume that there exists an exact solution $Q^{\ast }\in
G\left( R/5,\Phi ^{\ast },\Psi ^{\ast }\right) $ of the problem (\ref%
{3.13}), (\ref{4.1}) and that all conditions of Theorem \ref{th:2} hold.
Then for all $\delta \in \left( 0,\delta _{0}\right) $ the following
estimate holds 
\begin{equation}
\left\Vert Q_{n+1}-Q^{\ast }\right\Vert _{H^{1}\left( \Omega _{a}\right)
}\leq \theta ^{n}\left\Vert Q_{1}-Q_{\min }\right\Vert _{H^{3}\left( \Omega
\right) }+C_{1}\delta ^{\gamma },n=1,2,..  \label{5.11}
\end{equation}

\end{theorem}

\textbf{Proof. }Consider the nonlinear operator\textbf{\ }$Z\left( Q\right)
=\Delta Q+F\left( \nabla Q\right) $ in (\ref{3.13}). Then $Z:G\left( R,\Phi
,\Psi \right) \rightarrow L_{2}^{\lambda ,\nu _{0}}\left( \Omega \right) ,$
where the space $L_{2}^{\lambda ,\nu _{0}}\left( \Omega \right) $ is defined
as the space of vector functions $f\left( x\right) =\left(
f_{0},...,f_{N-1}\right) \left( x\right) ,x\in \Omega $ such that%
\begin{equation*}
L_{2}^{\lambda ,\nu _{0}}\left( \Omega \right) =\left\{ f:\left\Vert
f\right\Vert _{L_{2}^{\lambda ,\nu _{0}}\left( \Omega \right)
}^{2}=\int\limits_{\Omega }f^{2}\varphi _{\lambda ,\nu _{0}}^{2}dx<\infty
\right\} .
\end{equation*}%
Hence, the Fr\'{e}chet derivative $Z^{\prime }\left( Q\right) :H^{3}\left(
\Omega \right) \rightarrow L_{2}^{\lambda ,\nu _{0}}\left( \Omega \right) $
at any point $Q\in G\left( R,\Phi ,\Psi \right) $ acting on an element $%
h\left( x\right) =\left( h_{0},h_{1},...,h_{N-1}\right) \left( x\right) \in
H_{0}^{3}\left( \Omega \right) $ is defined as $Z^{\prime }\left( Q\right)
h=\Delta h+F^{\prime }\left( \nabla Q\right) \nabla h$. Let $\mathcal{L}%
\left( H_{0}^{3}\left( \Omega \right) ,L_{2}^{\lambda ,\nu _{0}}\left(
\Omega \right) \right) $ be the space of bounded linear operators mapping $%
H_{0}^{3}\left( \Omega \right) $ in $L_{2}^{\lambda ,\nu _{0}}\left( \Omega
\right) .$ It follows from results of section 8.2 of the book \cite{BKS}
that in order to prove this theorem, we should prove first that norms $%
\left\Vert Z^{\prime }\left( Q\right) \right\Vert _{L\left( H_{0}^{3}\left(
\Omega \right) ,L_{2}^{\lambda ,\nu _{0}}\left( \Omega \right) \right) }$
are uniformly bounded for $Q\in G\left( R,\Phi ,\Psi \right) .$ Second, we
should prove that the map 
\begin{equation}
Z^{\prime }\left( Q\right) :G\left( R,\Phi ,\Psi \right) \rightarrow 
\mathcal{L}\left( H_{0}^{3}\left( \Omega \right) ,L_{2}^{\lambda ,\nu
_{0}}\left( \Omega \right) \right)  \label{5.20}
\end{equation}%
is Lipschitz continuous on the set $G\left( R,\Phi ,\Psi \right) $. It
follows from the above that 
\begin{equation*}
\left\Vert Z^{\prime }\left( Q\right) h\right\Vert _{L_{2}^{\lambda ,\nu
_{0}}\left( \Omega \right) }\leq C_{1}e^{\lambda \xi ^{-\nu _{0}}}\left\Vert
h\right\Vert _{H^{3}\left( \Omega \right) },\forall h\in H_{0}^{3}\left(
\Omega \right) ,\forall Q\in G\left( R,\Phi ,\Psi \right) .
\end{equation*}%
Hence, 
\begin{equation}
\left\Vert Z^{\prime }\left( Q\right) \right\Vert _{\mathcal{L}\left(
H_{0}^{3}\left( \Omega \right) ,L_{2}^{\lambda ,\nu _{0}}\left( \Omega
\right) \right) }\leq C_{1}e^{\lambda \xi ^{-\nu _{0}}},\forall Q\in G\left(
R,\Phi ,\Psi \right) .  \label{5.3}
\end{equation}

To prove the Lipschitz continuity of the map (\ref{5.20}), we need to
estimate the norm $\left\Vert Z^{\prime }\left( Q_{1}\right) h-Z^{\prime
}\left( Q_{2}\right) h\right\Vert _{L_{2}^{\lambda ,\nu _{0}}\left( \Omega
\right) }$, for $Q_{1},Q_{2}\in G\left( R,\Phi ,\Psi \right) ,\ h\in
H_{0}^{3}\left( \Omega \right) .$ We have 
\begin{equation*}
Z^{\prime }\left( Q_{1}\right) h-Z^{\prime }\left( Q_{2}\right) h=\left[
F^{\prime }\left( \nabla Q_{1}\right) -F^{\prime }\left( \nabla Q_{2}\right) %
\right] \nabla h.
\end{equation*}%
Since $F\in C^{\infty }\left( \mathbb{R}^{3N}\right) ,$ then $\left\vert %
\left[ F^{\prime }\left( \nabla Q_{1}\right) -F^{\prime }\left( \nabla
Q_{2}\right) \right] \nabla h\right\vert \left( x\right) \leq
C_{1}\left\Vert Q_{1}-Q_{2}\right\Vert _{H^{3}\left( \Omega \right)
}\left\vert \nabla h\right\vert \left( x\right) $, $\forall Q_{1},Q_{2}\in
G\left( R,\Phi ,\Psi \right) ,\forall h\in H_{0}^{3}\left( \Omega \right)
,\forall x\in \overline{\Omega }.$ Hence, 
\begin{equation}
\left\Vert Z^{\prime }\left( Q_{1}\right) h-Z^{\prime }\left( Q_{2}\right)
h\right\Vert _{L_{2}^{\lambda ,\nu _{0}}\left( \Omega \right) }\leq
C_{1}e^{\lambda \xi ^{-\nu _{0}}}\left\Vert Q_{1}-Q_{2}\right\Vert
_{H^{3}\left( \Omega \right) }\left\Vert h\right\Vert _{H^{3}\left( \Omega
\right) },  \label{5.4}
\end{equation}%
which proves the Lipschitz continuity of the map (\ref{5.20}). Hence,
Theorem \ref{th:1}, (\ref{5.3}), (\ref{5.4}) and the result of section 8.2
of the book \cite{BKS} guarantee that (\ref{5.2}) holds, as long as 
\begin{equation}
Q_{n}\in G\left( R,\Phi ,\Psi \right) ,\forall n=1,2,...  \label{5.5}
\end{equation}

We show now that (\ref{5.5}) is true. Since $Q_{1},Q_{\min }$ $\in G\left(
R/4,\Phi ,\Psi \right) ,$ then 
\begin{equation}
\left\Vert Q_{1}-Q_{\min }\right\Vert _{H^{3}\left( \Omega \right) }<R/2.
\label{5.50}
\end{equation}%
It follows from (\ref{4.62}) that norms $\Vert J_{\lambda ,\alpha }^{\prime
}\left( Q\right) \Vert _{H^{3}\left( \Omega \right) }$ are uniformly bounded
for all $Q\in G\left( R,\Phi ,\Psi \right) .$ Choose a sufficiently small
number $\beta =\beta \left( \lambda ,\alpha ,G\left( R,\Phi ,\Psi \right)
\right) \in \left( 0,1\right) $ such that 
\begin{equation}
\beta \Vert J_{\lambda ,\alpha }^{\prime }\left( Q\right) \Vert
_{H^{3}\left( \Omega \right) }<R/4,\ \forall Q\in G\left( R,\Phi ,\Psi
\right) .  \label{5.6}
\end{equation}%
Hence, (\ref{5.1}) and (\ref{5.6}) imply that $\left\Vert Q_{2}\right\Vert
_{H^{3}\left( \Omega \right) }\leq R/4+R/4=R/2.$ Hence, $Q_{2}\in G\left(
R,\Phi ,\Psi \right) .$ Thus, (\ref{5.2}) is true for $n=1$. By (\ref{5.2})
and (\ref{5.50}) $\left\Vert Q_{2}-Q_{\min }\right\Vert _{H^{3}\left( \Omega
\right) }\leq \theta R/2.$ Next, $\left\Vert Q_{2}\right\Vert _{H^{3}\left(
\Omega \right) }\leq \left\Vert Q_{2}-Q_{\min }\right\Vert _{H^{3}\left(
\Omega \right) }+\left\Vert Q_{\min }\right\Vert _{H^{3}\left( \Omega
\right) }.$ Hence, 
\begin{equation*}
\left\Vert Q_{2}\right\Vert _{H^{3}\left( \Omega \right) }\leq \left\Vert
Q_{\min }\right\Vert _{H^{3}\left( \Omega \right) }+\frac{R}{2}\theta
<\left( \frac{1}{4}+\frac{\theta }{2}\right) R<\frac{3}{4}R.
\end{equation*}%
Suppose that $Q_{2},...,Q_{n}\in G\left( 3R/4,\Phi ,\Psi \right) $. Then (%
\ref{5.2}) holds for these terms. By (\ref{5.1}) and (\ref{5.6}) $\left\Vert
Q_{n+1}\right\Vert _{H^{3}\left( \Omega \right) }<3R/4+R/4=R.$ Hence, $%
Q_{n+1}\in G\left( R,\Phi ,\Psi \right) .$ Hence, by (\ref{5.50}) and the
triangle inequality%
\begin{equation*}
\left\Vert Q_{n+1}\right\Vert _{H^{3}\left( \Omega \right) }\leq \left\Vert
Q_{\min }\right\Vert _{H^{3}\left( \Omega \right) }+\frac{R}{2}\theta
^{n}<\left( \frac{1}{4}+\frac{\theta ^{n}}{2}\right) R<\frac{3}{4}R.
\end{equation*}%
Hence, $Q_{n+1}\in G\left( 3R/4,\Phi ,\Psi \right) $. Thus, we have
established that (\ref{5.5}) holds, which, in turn implies (\ref{5.2}).

Now we prove (\ref{5.11}). Using (\ref{5.10}), (\ref{5.2}) and the triangle
inequality, we obtain%
\begin{eqnarray*}
&&\theta ^{n}\left\Vert Q_{1}-Q_{\min }\right\Vert _{H^{3}\left( \Omega
\right) } \geq \left\Vert Q_{n+1}-Q_{\min }\right\Vert _{H^{3}\left( \Omega
\right) }\geq \left\Vert Q_{n+1}-Q_{\min }\right\Vert _{H^{1}\left( \Omega
_{a}\right) } \\
&&\geq \left\Vert Q_{n+1}-Q^{\ast }\right\Vert _{H^{1}\left( \Omega
_{a}\right) }-\left\Vert Q^{\ast }-Q_{\min }\right\Vert _{H^{1}\left( \Omega
_{a}\right) }\geq \left\Vert Q_{n+1}-Q^{\ast }\right\Vert _{H^{1}\left(
\Omega _{a}\right) }-C_{1}\delta ^{\gamma }.
\end{eqnarray*}%
Thus, $\left\Vert Q_{n+1}-Q^{\ast }\right\Vert _{H^{1}\left( \Omega
_{a}\right) }\leq \theta ^{n}\left\Vert Q_{1}-Q_{\min }\right\Vert
_{H^{3}\left( \Omega \right) }+C_{1}\delta ^{\gamma },$ which is (\ref{5.11}%
). $\hfill\square $

\section{The 1-d coefficient inverse problem}

\label{sec:1dprob}

In this section we consider the 1-d analog of the above CIP. Our motivation
for considering this case is twofold. First, in some practical cases, only
1-d data are available, see, e.g., \cite{KBKSNF} for experimental data
measured by US Army Research Laboratory for devices mimicking explosives,
which is our target application. Second, since numerical computation of the
1-d problem is simple and fast, we can use it to quickly analyze influence
of different parameters on the performance of the algorithm in order to
choose optimal ones which may be used for the 3-d case as well. Analytical
results for our 1-d CIP are quite similar to the 3-d case. Therefore, we
only briefly state them below.

The forward problem in the 1-d case is 
\begin{eqnarray}
&&c\left( x\right) u_{tt}=u_{xx}+\delta (x-x^{0})f(t),\ \left( x,t\right)
\in \mathbb{R}\times \left( 0,\infty \right) ,  \label{6.1} \\
&&u\left( x,0\right) =0,u_{t}\left( x,0\right) =0.  \label{6.2}
\end{eqnarray}%
For simplicity, we choose $\Omega =\left( 0,b\right) $, where $b>0$ is a
constant, and the function $c\left( x\right) $ satisfies the following
analogs of conditions (\ref{2.1}) 
\begin{equation}
c\in C^{1+\alpha }\left( \mathbb{R}\right) ,0<c_{0}\leq c\left( x\right)
\leq 1+d,\forall x\in \mathbb{R},\ c\left( x\right) =1,x\notin \left(
0,b\right) .  \label{6.3}
\end{equation}

\textbf{CIP 2. }\emph{Suppose that in (\ref{6.1}) the source location }$%
x^{0}<0$\emph{. Determine the function }$c(x)$\emph{\ for }$x\in \left(
0,b\right) ,$\emph{\ assuming that the following functions }$p_{1}\left(
t\right) ,p_{2}\left( t\right) $\emph{\ are given}%
\begin{equation}
u\left( 0,t\right) =p_{1}\left( t\right) ,u_{x}\left( 0,t\right)
=p_{2}\left( t\right) ,t\in \left( 0,\infty \right) .  \label{6.5}
\end{equation}

The CWF in the 1-d case can be chosen as $\varphi _{\lambda }\left( x\right)
=e^{-\lambda x}$ which is different from the one in (\ref{4.2}). Note that
this CWF makes numerical computation more efficient. To justify the use of
this CWF, we use the Carleman estimate of Lemma \ref{le:2}. We omit its
proof, since it can be obtained via a slight modification of arguments on
pages 188, 189 of \cite{KT}.

\begin{lemma}\label{le:2}
The following Carleman estimate holds for
all functions $u\in H^{2}\left( 0,b\right) $ with a number $%
C_{2}=C_{2}\left( b\right) >0$ depending only on $b$ and for
all $\lambda >1$ 
\begin{equation*}
\begin{split}
\int\limits_{0}^{b}\left( u^{\prime \prime }\right) ^{2}\varphi _{\lambda
}^{2}\left( x\right) dx& +C_{2}\left[ \lambda \left( u^{\prime }\left(
0\right) \right) ^{2}+\lambda ^{3}u^{2}\left( 0\right) \right] \\
& \geq C_{2}\int\limits_{0}^{b}\left[ \left( u^{\prime \prime }\right)
^{2}+\lambda \left( u^{\prime }\right) ^{2}+\lambda ^{3}u^{2}\right] \varphi
_{\lambda }^{2}\left( x\right) dx,
\end{split}%
\end{equation*}%
which implies that %
\begin{equation}
\int\limits_{0}^{b}\left( u^{\prime \prime }\right) ^{2}\varphi _{\lambda
}^{2}\left( x\right) dx+C_{2}\left[ \lambda \left( u^{\prime }\left(
0\right) \right) ^{2}+\lambda ^{3}u^{2}\left( 0\right) \right] \geq
C_{2}e^{-2\lambda b}\left\Vert u\right\Vert _{H^{2}\left( 0,b\right) }^{2}.
\label{6.6}
\end{equation}

\end{lemma}


The one-dimensional analog of the set $G=G\left( R,\Phi ,\Psi \right) $ in (%
\ref{4.3}) is 
\begin{equation*}
G_{1}=G_{1}\left( R,\Phi _{0},\Psi _{0},\Psi _{b}\right) =\left\{ 
\begin{array}{c}
Q=\left( q_{0},...,q_{N-1}\right) ^{T}\in H^{2}\left( 0,b\right) :\left\Vert
Q\right\Vert _{H^{2}\left( 0,b\right) }<R, \\ 
Q\left( 0\right) =\Phi _{0},\ Q^{\prime }\left( 0\right) =\Psi _{0},\
Q^{\prime }(b)=\Psi _{b},%
\end{array}%
\right.
\end{equation*}%
where vectors $\Phi _{0}=(\Phi _{0,0}, \Phi_{0,1},\dots ,\Phi _{0,N-1}),\
\Psi _{0}=(\Psi _{0,0},\Psi_{0,1},\dots ,\Psi _{0,N-1})$ and $\Psi
_{b}=(\Psi _{b,0},\Psi_{0,1},\dots ,\Psi _{b,N-1})$ belong to $\mathbb{R}%
^{N}.$ Here the Neumann boundary data $\Psi _{b}$ at $x=b$ is derived from
the Absorbing Boundary Condition (ABC) \cite{EM:PNAS1977}, keeping in mind
that $u(x,t)$ is an out-going wave at $x=b$, see section \ref{subsec:numfp}.
By Embedding Theorem, $H^{2}\left( 0,b\right) \subset C^{1}\left[ 0,b\right] 
$ and $\left\Vert Q\right\Vert _{C^{1}\left[ 0,b\right] }<CR,\forall Q\in
G_{1}.$

It follows from (\ref{6.6}) that there is no need to use the regularization
term in the 1-d version of the functional (\ref{4.5}). Thus, we use in our
numerical study the following analog of $J_{\lambda ,\alpha }$%
\begin{equation}
\overline{J}_{\lambda }\left( Q\right) =\int\limits_{0}^{b}\left[ Q^{\prime
\prime }+F\left( Q^{\prime }\right) \right] ^{2}\varphi _{\lambda
}^{2}\left( x\right) dx.  \label{6.9}
\end{equation}

Theorem \ref{th:4} is the 1-d analog of Theorem \ref{th:1}, and the proof is
similar.

\begin{theorem}\label{th:4}
There exists a sufficiently large number %
$\lambda _{2}=\lambda _{2}\left( G_{1},F\right) >1$ depending only on 
$G_{1}$ and $F$ such that for all $\lambda \geq \lambda
_{2} $ the functional $\overline{J}_{\lambda }\left( Q\right) $
 is strictly convex on the set $G_{1}$, i.e., there exists a
constant $C_{3}=C_{3}\left( G_{1},F\right) >0$ depending only on $%
G_{1}$ and $F$ such that 
\begin{equation}
\overline{J}_{\lambda }\left( Q_{2}\right) -\overline{J}_{\lambda }\left(
Q_{1}\right) -\overline{J}_{\lambda }^{\prime }\left( Q_{1}\right) \left(
Q_{2}-Q_{1}\right) \geq C_{3}e^{-2\lambda b}\left\Vert
Q_{2}-Q_{1}\right\Vert _{H^{2}\left( 0,b\right) }^{2},  \label{6.10}
\end{equation}%
where $\overline{J}_{\lambda }^{\prime }\left( Q_{1}\right) $
is the Fr\'{e}chet derivative of the functional $\overline{J}_{\lambda }$%
 at the point $Q_{1}.$

\end{theorem}


\begin{remark}
\label{re:2}
\end{remark}


1. Although we do not need the knowledge of the vector $\Psi _{b}$ in the
proof of this theorem, we use this knowledge for a better stability of our
numerical method. The strict convexity constant $C_{2}e^{-2\lambda b}$ in (%
\ref{6.10}) is small either for large $\lambda $ or for large $b$.
Therefore, it is expected that the functional $\overline{J}_{\lambda }$ is
more sensitive to the change of $Q$ near the point $\left\{ x=0\right\} $
than at points far from it. In other words, the slope of $\overline{J}%
_{\lambda }$ should be large near $\left\{ x=0\right\} $ and small far away
from $\left\{ x=0\right\} $. Consequently, it may be hard to obtain accurate
approximation of the solution far away from $x=0$. To remedy this, some sort
of the layer stripping procedure may be used. The idea of this layer
stripping procedure is that we first consider the integral over $\left(
0,b_{1}\right) $ for a small value of $b_{1}\in \left( 0,b\right) $. Next,
we consider the integral over $\left( b_{1},2b_{1}\right) ,$ etc. A balance
between values of $\lambda $ and $b_{1}$ should be found in numerical
experiments.

2. Another option for enhancing the accuracy of the computed coefficient,
which we use here, is to refine it via a gradient-based optimization method
for the original time domain problem, see section \ref{subsec:localmethod}.
This results in a two-stage numerical procedure, see Chapters 4,5 of \cite%
{BK} for the idea of such a procedure for a different numerical method.\
More precisely: (1) on the first stage a globally convergent numerical
method addresses the most difficult question of obtaining a point in a small
neighborhood of the exact coefficient without any advanced knowledge of that
neighborhood, and (2) on the second stage a locally convergent numerical
method refines the solution of the first stage via starting its iteration
from that point.

\section{Numerical implementation}

\label{sec:6}

In this section we describe details of our numerical implementation of the
proposed algorithm. We also test a two-stage numerical procedure mentioned
in item 2 of Remark \ref{re:2}.

\subsection{Solving the forward problem (\protect\ref{6.1})--(\protect\ref%
{6.2})}

\label{subsec:numfp} In numerical computation, we solve the forward problem (%
\ref{6.1})--(\ref{6.2}) in a bounded interval $(k,d)$ such that $k<0<b<g$.
Recall that $c(x)=1,\ x\notin (0,b)$. We consider the case that $x^{0}\leq k$
and rewrite $u(x,t)=u^{i}(x,t)+u^{s}(x,t)$, where $u^{i}$ is the incident
wave and $u^{s}$ is the scattered wave. The incident wave $u^{i}$ satisfies (%
\ref{6.1})--(\ref{6.2}) with $c(x)\equiv 1$ and is given by the following
formula 
\begin{equation*}
u^{i}(x,t)=%
\begin{cases}
f(t-|x-x^{0}|), & t\geq |x-x^{0}|, \\ 
0, & 0\leq t<|x-x^{0}|.%
\end{cases}%
\end{equation*}%
To approximate the wave propagation in the whole 1-d space $\mathbb{R}$ by
the problem in the bounded interval $(k,d)$, we keep in mind the fact that
the scattered wave is out-going in both directions.\ Thus, we assume that
the function $u^{s}(x,t)$ satisfies the ABC at $x=k$ and $x=g$. This means
that we solve the following problem for $u^{s}$: 
\begin{eqnarray}
&&c(x)u_{tt}^{s}(x,t)-u_{xx}^{s}(x,t)=[1-c(x)]u_{tt}^{i}(x,t),(x,t)\in
(k,d)\times (0,T),  \label{eq:numfp1} \\
&&u^{s}(x,0)=u_{t}^{s}(x,0)=0,\ x\in (k,d),  \label{eq:numfp2} \\
&&u_{x}^{s}(k,t)=u_{t}^{s}(k,t),\ u_{x}^{s}(g,t)=-u_{t}^{s}(g,t),\ t\in
(0,T).  \label{eq:numfp3}
\end{eqnarray}

In the numerical examples presented below, we choose $k=x^{0}=-0.2$, $b=0.4$%
, $g=0.5$ and $T=2$. The waveform $f(t)$ of the incident wave is chosen to
be 
\begin{equation*}
f(t)=A(t-0.2)e^{-\omega ^{2}(t-0.2)^{2}},
\end{equation*}%
where $\omega =30$ and $A=\sqrt{2}\omega e^{1/2}$. The constant $A$ is used
as the normalization factor. The problem (\ref{eq:numfp1})--(\ref{eq:numfp3}%
) is solved by an explicit Finite Difference scheme with uniform grids in
both $x$ and $t$ with step sizes of $\Delta x=0.005$ and $\Delta t=0.001$.
This results in $141$ grid points in space and $2001$ points in time.

In order to simulate noisy measurements, we add additive noise of $10\%$ (in
the $L_2$ norm) to the simulated data.

\subsection{Discretization of the objective functional (\protect\ref{6.9})}

Consider a partition of the interval $(0,b)$ into $M$ sub-intervals by the
grid points $0=x_{0}<x_{1}<\cdots <x_{M}=b$ with $x_{i+1}-x_{i}=h$, $%
i=0,\dots ,M-1$. We define the discrete unknown function $%
Q_{h}:=\{q_{j}^{i},\ j=0,\dots ,N-1,\ i=0\dots ,M\}$ with $%
q_{j}^{i}=q_{j}(x_{i})$. We approximate the functional (\ref{6.9}) by the
following discrete version using a forward Finite Difference scheme 
\begin{equation}
\bar{J}_{h,\lambda }(Q_{h}):=h\sum\limits_{j=0}^{N-1}\sum\limits_{i=1}^{M-1}%
\left[ J_{j}^{i}(Q_{h})\right] ^{2}\varphi _{\lambda }^{2}(x_{i}),
\label{eq:numqn}
\end{equation}%
where 
\begin{equation}
J_{j}^{i}(Q_{h})=\frac{q_{j}^{i+1}-2q_{j}^{i}+q_{j}^{i-1}}{h^{2}}%
+\sum\limits_{m=0}^{N-1}\sum\limits_{n=0}^{N-1}F_{jmn}\frac{%
(q_{m}^{i+1}-q_{m}^{i})(q_{n}^{i+1}-q_{n}^{i})}{h^{2}},  \label{eq:numqn2}
\end{equation}%
for $j=0,\dots ,N-1$ and $i=1,\dots ,M-1$. Note that from the boundary
conditions for $Q$, we have 
\begin{equation}
q_{j}^{0}=\Phi _{0,j},\quad q_{j}^{1}\approx q_{j}^{0}+h\Psi _{0,j},\quad
q_{j}^{M}\approx q_{j}^{M-1}+h\Psi _{b,j}.  \label{eq:numqn3}
\end{equation}%
Hence, the unknowns to be determined are $q_{j}^{i}$, $j=0,\dots ,N-1$, $%
i=2,\dots ,M-1$. The gradient of the functional $\bar{J}_{h,\lambda }$ can
be easily derived from (\ref{eq:numqn}) and (\ref{eq:numqn2}). The first
guesses $q_{j,init}^{i}$ for $q_{j}^{i}$ in minimizing the functional (\ref%
{eq:numqn}) are chosen as 
\begin{equation}
q_{j,init}^{i}=\Phi _{0,j},\ j=0,\dots ,N-1,\ i=2,\dots ,M-1.
\label{eq:qn_init}
\end{equation}

We do not have a proof of the global strict convexity of the discrete
functional (\ref{eq:numqn}) at the moment and leave this for future work.

We should mention that the spatial mesh size $\Delta x$ for solving the
forward problem (\ref{eq:numfp1})--(\ref{eq:numfp3}) is chosen small enough,
which should be about $1/10$ of the wavelength of the incident wave, for the
accuracy of the Finite Difference scheme. However, our numerical computation
have indicated that in order to enhance the stability of the minimization of
the functional (\ref{eq:numqn}), the grid size $h$ in the discretization of
the function $Q$ should not be chosen too small. In the tests below, $h$ is
chosen to be $h=0.025$. After getting the values of the coefficient $c(x)$
at these grid points, we linearly interpolate it to the finer grid of size $%
\Delta x$ for a local method presented below.

Despite the fact that the functional (\ref{6.9}) is strictly convex on the
bounded set $G_{1}$, our numerical tests work even without any constraints
applied to the unknown coefficients. That means that we only need to solve
unconstrained optimization problems.

\subsection{Combination of the proposed globally convergent method with a
locally convergent method}

\label{subsec:localmethod}

Although at least the continuous counterpart (\ref{6.9}) of the functional (%
\ref{eq:numqn}) is guaranteed to be strictly convex on the set $G_{1}$ for $%
\lambda $ large enough, we have observed numerically that the slope of $\bar{%
J}_{h,\lambda }$ is quite small, especially with respect to large values of $%
i$, i.e., for points far away from the location of measurement. Due to this
reason, it is hard to obtain accurate results using the globally convergent
method alone. In order to enhance the accuracy of the computed coefficient
obtained by the proposed globally convergent method, we combine it with a
locally convergent gradient-based method, which uses the result of the
former as an initial guess. Given the boundary data (\ref{6.5}), the forward
problem (\ref{eq:numfp1})--(\ref{eq:numfp3}) can be replaced with the
following one: 
\begin{eqnarray}
&&c(x)u_{tt}^{s}(x,t)-u_{xx}^{s}(x,t)=[1-c(x)]u_{tt}^{i}(x,t),(x,t)\in
(0,d)\times (0,T),  \notag \\
&&u^{s}(x,0)=u_{t}^{s}(x,0)=0,\ x\in (0,d),  \notag \\
&&u_{x}^{s}(0,t)=p_{2}(t)-u_{x}^{i}(0,t),\ t\in (0,T),  \notag \\
&&u_{x}^{s}(g,t)=-u_{t}^{s}(g,t),\ t\in (0,T).  \notag
\end{eqnarray}%
The coefficient $c(x),\ x\in (0,b)$, is determined by minimizing the
following objective functional 
\begin{equation}  \label{eq:local1}
M_{\alpha }(c):=\frac{1}{2}\int\limits_{0}^{T}[u(0,t;c)-p_{1}(t)]^{2}dt+%
\frac{1}{2}\alpha \mathcal{R}(c-c_{glob}),
\end{equation}%
where $c_{glob}$ is the coefficient computed by the globally convergent
method and $\alpha \mathcal{R}(c-c_{glob})$, $\alpha >0$, is a Tikhonov-type
regularization term. In our tests, we choose $\mathcal{R}(c-c_{glob})=\Vert
c-c_{glob}\Vert _{H^{1}(0,b)}$. Using the adjoint equation method, it is
straightforward to derive the following formula for the gradient $M_{\alpha
}^{\prime }(c)$ of $M_{\alpha }(c)$: 
\begin{equation*}
M_{\alpha }^{\prime }(c)\left( x\right) =\int_{0}^{T}u_{t}(\cdot ,t)\eta
_{t}(\cdot ,t)dtdt+\frac{1}{2}\alpha \nabla \mathcal{R}(c-c_{glob}),
\end{equation*}%
where $\eta $ is the solution to the following adjoint problem 
\begin{eqnarray}
&&c(x)\eta _{tt}(x,t)-\eta _{xx}(x,t)=0,(x,t)\in (0,g)\times (0,T),  \notag
\\
&&\eta (x,T)=\eta _{t}(x,T)=0,\ x\in (0,g),  \notag \\
&&\eta _{x}(0,t)=p_{1}(0,t)-u(0,t),\ t\in (0,T),  \notag \\
&&\eta _{x}(g,t)=\eta _{t}(g,t),\ t\in (0,T).  \notag
\end{eqnarray}

In the following, we call the globally convergent method as Step 1 and the
locally convergent method as Step 2 of this hybrid algorithm. Furthermore,
we have observed in our numerical test that if the unknown coefficient is to
be determined in a large interval $(0,b)$, then the results of Step 2 may
not be very accurate. Therefore, we propose an additional step, Step 3, as
described in the following algorithm. %
%
%
%

\noindent\textbf{Algorithm:}

\begin{itemize}
\item Given the data: $p_1(t) = u(0,t)$, $p_2(t) = u_x(0,t)$, $t\in (0,T)$.
Choose $N$ Laguerre's coefficients.

\item Step 1 (the globally convergent method):

\begin{enumerate}
\item[1.1.] Compute the functions $\Phi _{0,k}$, $\Psi _{0,k}$ and $\Psi
_{b,k}$, $k=0,\dots ,N-1$.

\item[1.2.] Find a minimizer $\overline{Q}_{h}$ of the functional (\ref%
{eq:numqn}), starting from the initial guess (\ref{eq:qn_init}).

\item[1.3.] Compute the coefficient values $c_{glob}(x_{i}),\ i=0,\dots ,M,$
given the Laguerre's coefficients $\overline{Q}_{h}$.
\end{enumerate}

\item Step 2 (locally convergent method):

\begin{enumerate}
\item[2.1.] Linearly interpolate the coefficient $c_{glob}$ from the grid of
step size $h$ to the grid of step size $\Delta x$.

\item[2.2.] Compute $c_{local,1}(x),\ x\in (0,b)$, by minimizing functional (%
\ref{eq:local1}), starting from $c_{glob}$ as an initial guess.
\end{enumerate}

\item Step 3 (locally convergent method applied to a reduced spatial
interval):

\begin{enumerate}
\item[3.1.] Reduce the interval $(0,b)$ to $(0,b_{1})$, where $b_{1}$ is
determined as follows 
\begin{equation*}
b_{1}=\min \{x\in (0,b):|c_{local,1}(x)-1|\leq \epsilon \text{ and }%
\max_{y\in (0,x)}|c_{local,1}(y)-1|>\epsilon \}.
\end{equation*}

%

\item[3.2.] Compute an update $c_{local,2}$ of the coefficient by minimizing
the functional (\ref{eq:local1}) in $(0,b_{1})$, starting from $c_{local,1}$
as the initial guess.
\end{enumerate}
\end{itemize}

The choice of $b_1$ in Step 3.1 means that in Step 3 we refine the
coefficient value only in the interval closest to the measurement location
in which its value is substantially different from that of the background
medium. Here, $\epsilon $ is a truncation parameter which should be chosen
in numerical experiments.

The minimization problems on all steps of this algorithm are solved by the
Sequential Quadratic Programming method for unconstrained optimization
problems which is implemented in Matlab Optimization Toolbox.

\section{Numerical examples}

\label{sec:numexa}

In this section we present a limited testing of the above algorithm for some
numerically simulated data. We also compare its performance with the above
locally convergent method alone using the coefficient of the homogeneous
medium as the first guess. Numerical results for experimental data in both
1-d and 3-d cases are under consideration and will be reported in future
work.

\begin{figure}[tph]
\centering
\subfloat[]{\includegraphics[width=0.42\textwidth,height =
0.32\textwidth]{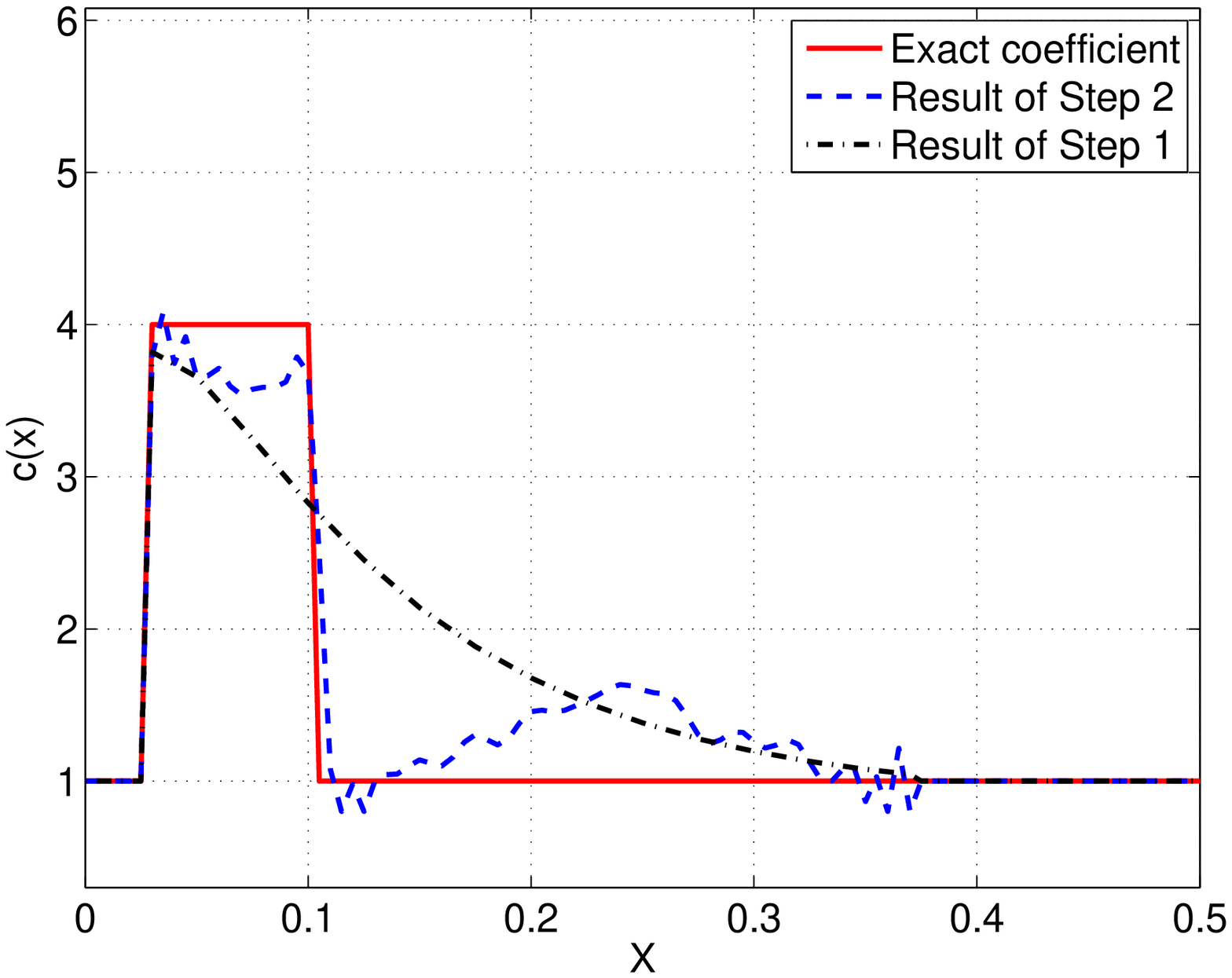}} 
\subfloat[]{\includegraphics[width=0.42\textwidth,height =
0.32\textwidth]{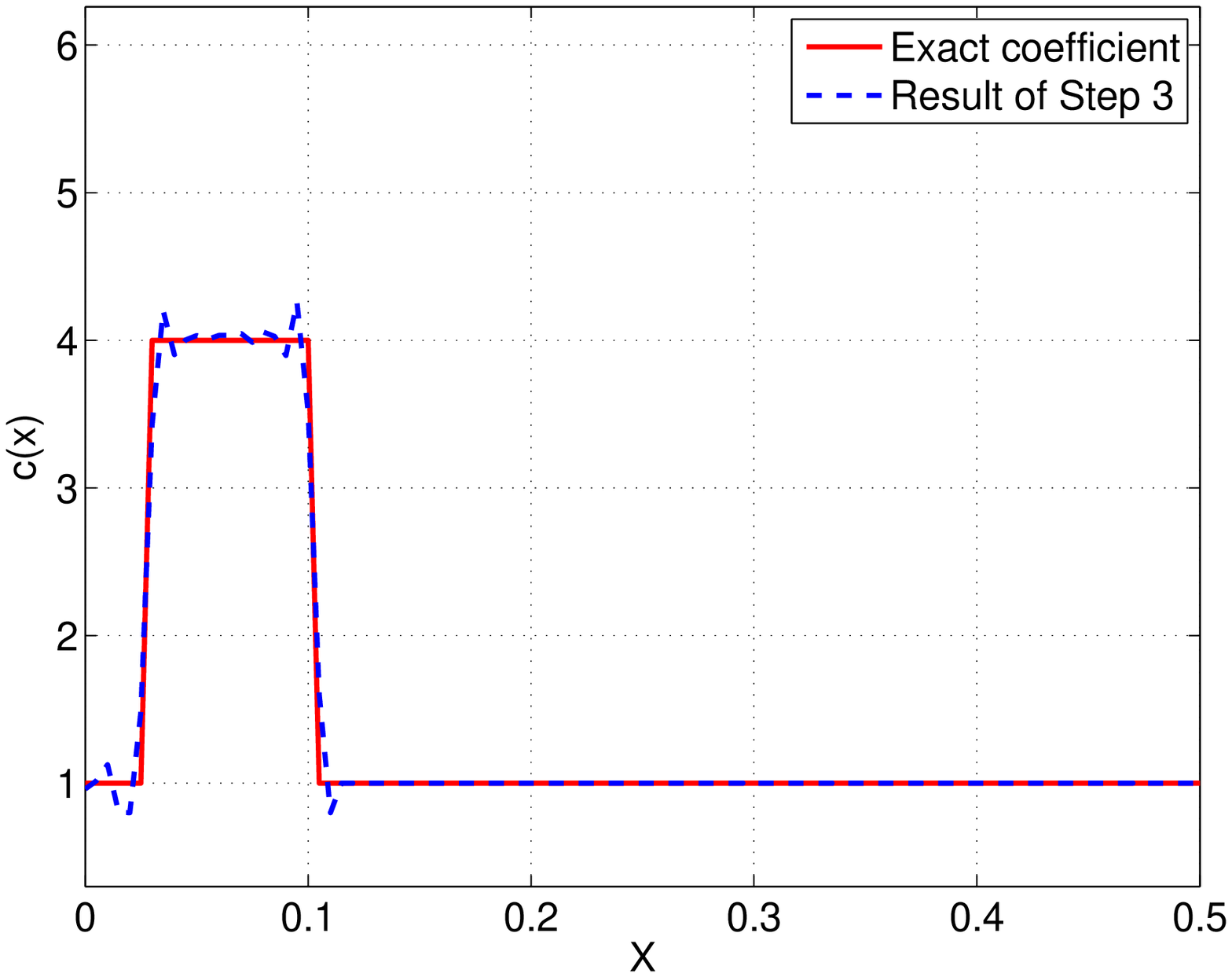}}
\par
\subfloat[]{\includegraphics[width=0.42\textwidth,height =
0.32\textwidth]{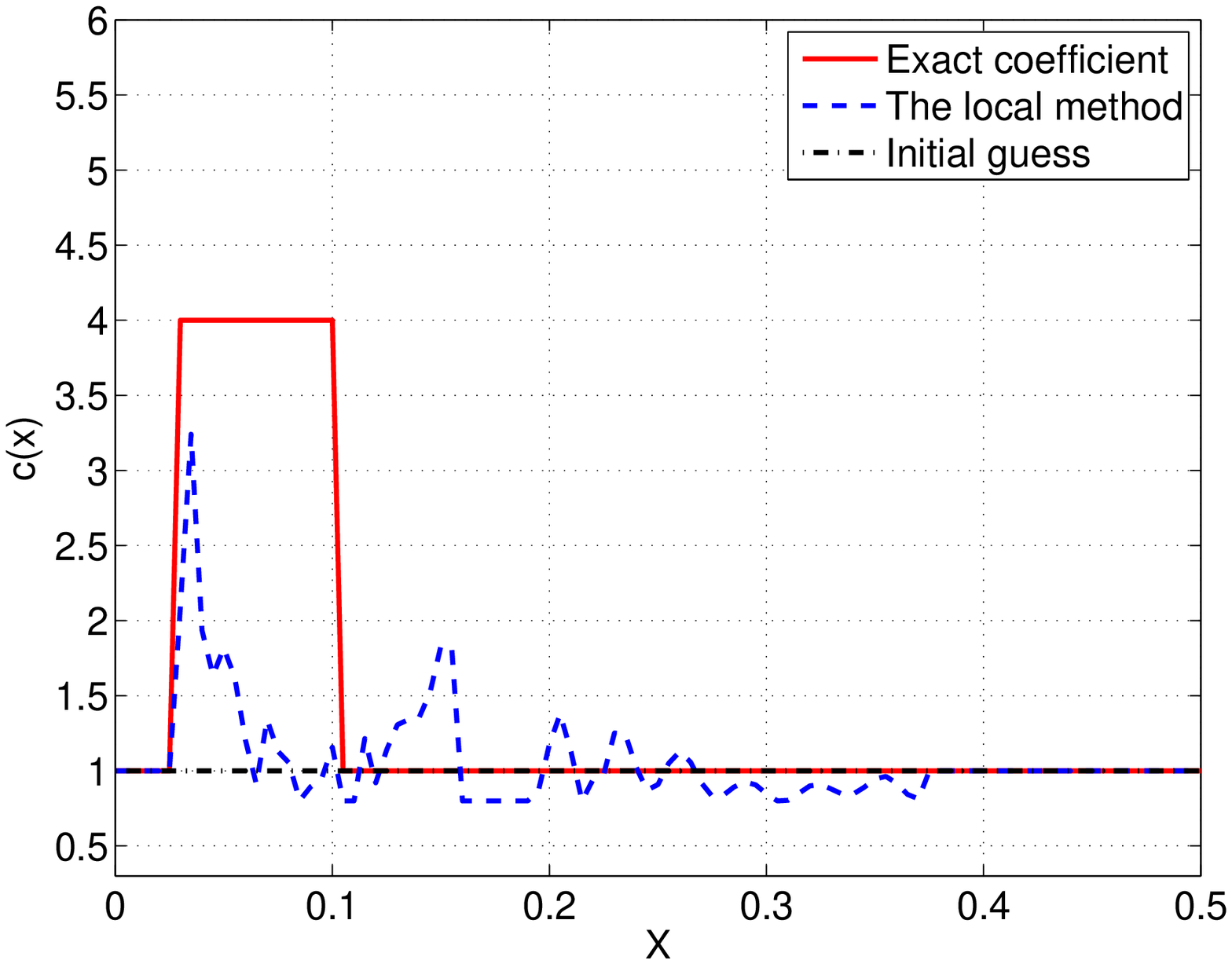}} 
\subfloat[]{\includegraphics[width=0.42\textwidth,height =
0.32\textwidth]{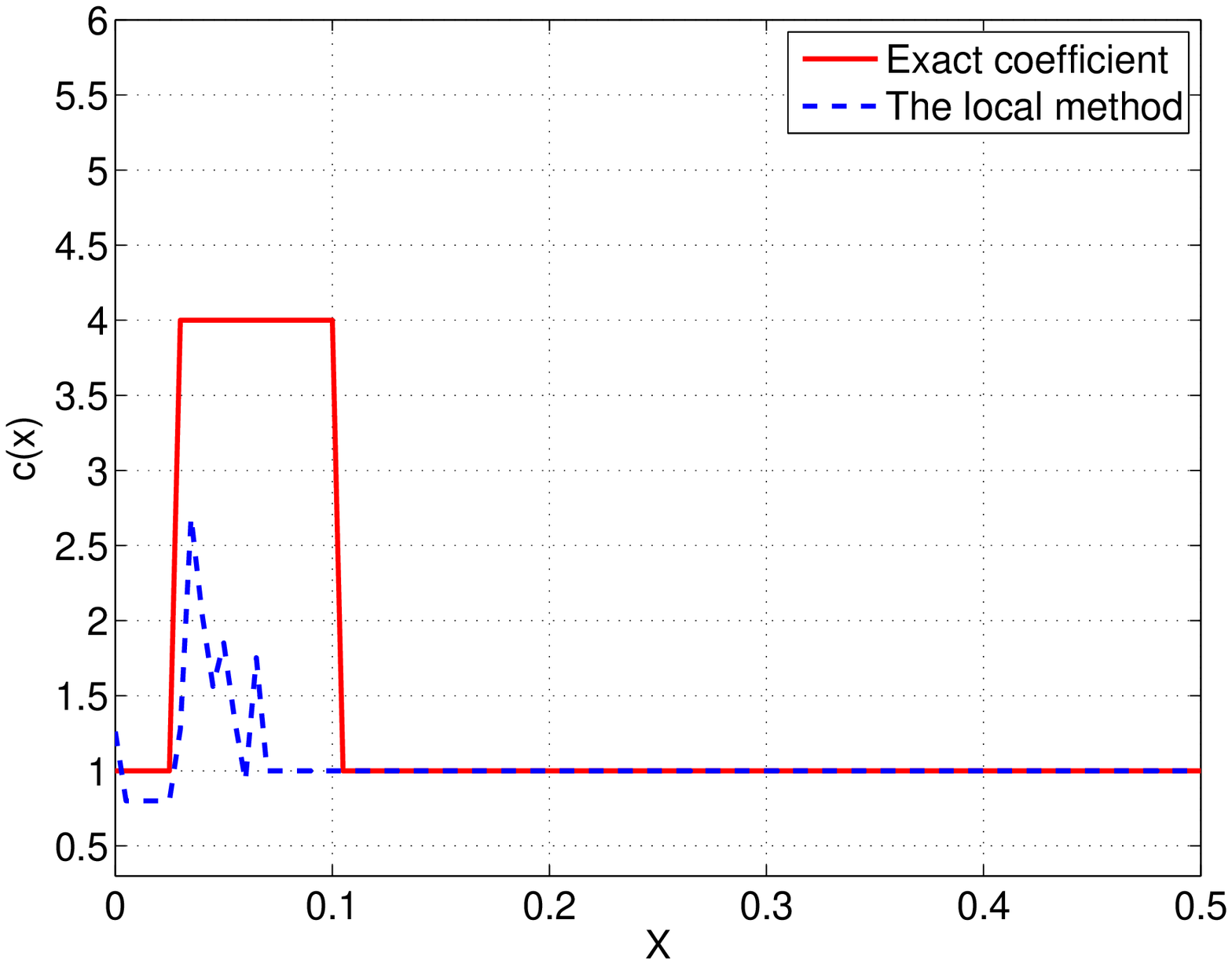}}
\caption{Reconstruction of the coefficient of Example 1. (a) Result of Steps
1 and 2, (b) Result of Step 3, (c) Result of Step 2 starting from the
homogeneous medium as the first guess, (d) Result of Step 3 applied to the
result of (c). }
\label{fig:exa1}
\end{figure}

Since our target application is in imaging of an abnormal object placed in a
homogeneous medium, we mainly test the proposed algorithm with a
discontinuous coefficient. The locations of the discontinuities represent
the location of the target. As mentioned in Remarks \ref{re:2}, it is hard
to obtain accurate reconstructions at locations far away from the
measurement point. However, this is not a serious restriction from the
practical standpoint. Indeed, our experience of working with 3-d time
resolved backscattering experimental data \cite{BNKF,NBKF,NBKF1} tells us
that, using the so-called data propagation procedure in data pre-processing,
we can approximate quite well both the distance from the measurement point
to the target and the Dirichlet and Neumann backscattering data near the
target. Thus, we assume below that the target is close to the measurement
point.

In the following examples, the parameters were chosen as follows: The pseudo
frequencies were $s\in \left[ 4,15\right] $ with the integration step size
in (\ref{eq:qn}) $\Delta s=0.05$. The number of Laguerre's functions was $%
N=11$. The coefficients $\lambda $ in the CWF was $\lambda =3$. The
regularization parameter $\alpha =0.001$ and the truncation value $\epsilon
=0.2$.

\begin{figure}[tph]
\centering
\subfloat[]{\includegraphics[width=0.42\textwidth,height =
0.32\textwidth]{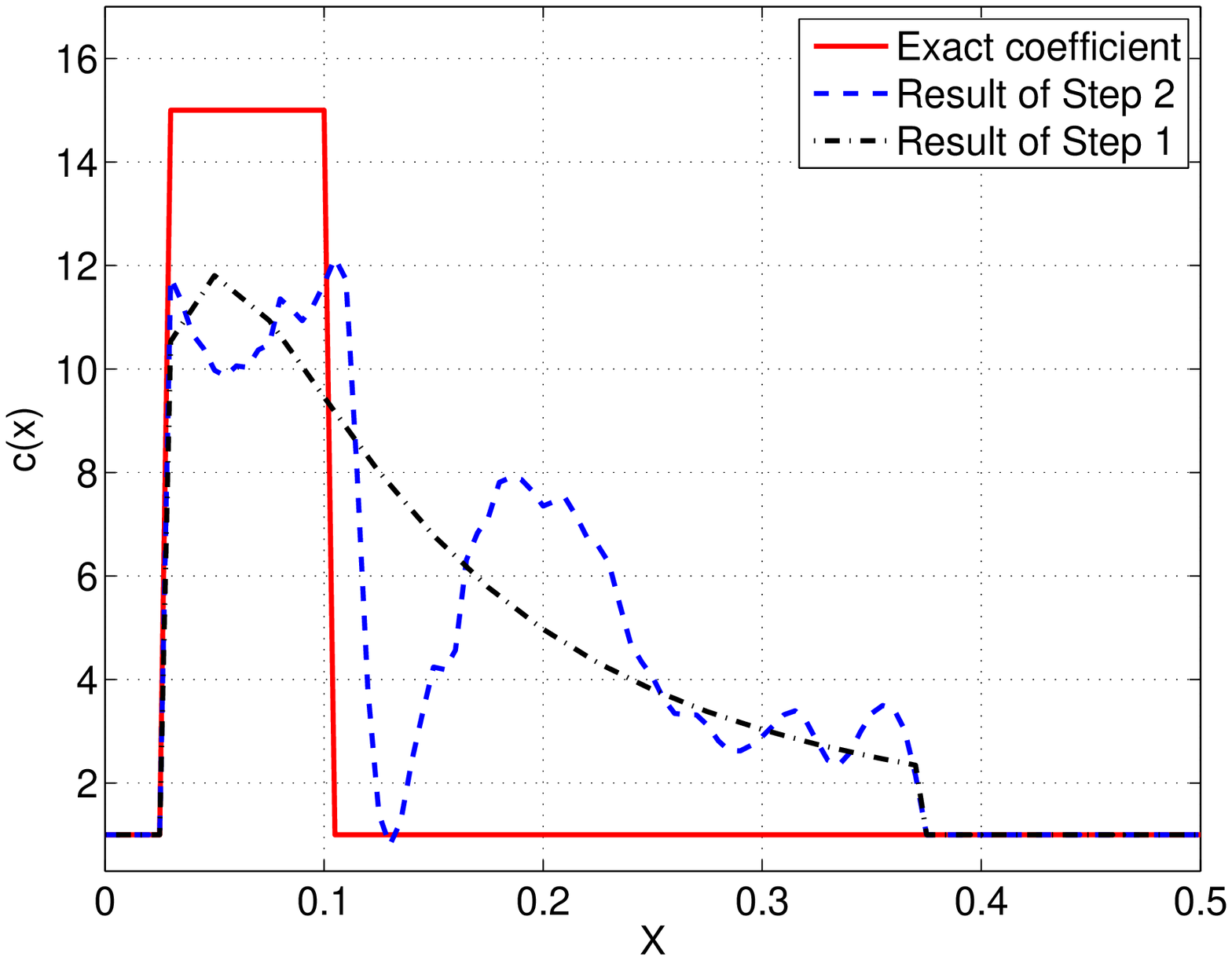}} 
\subfloat[]{\includegraphics[width=0.42\textwidth,height =
0.32\textwidth]{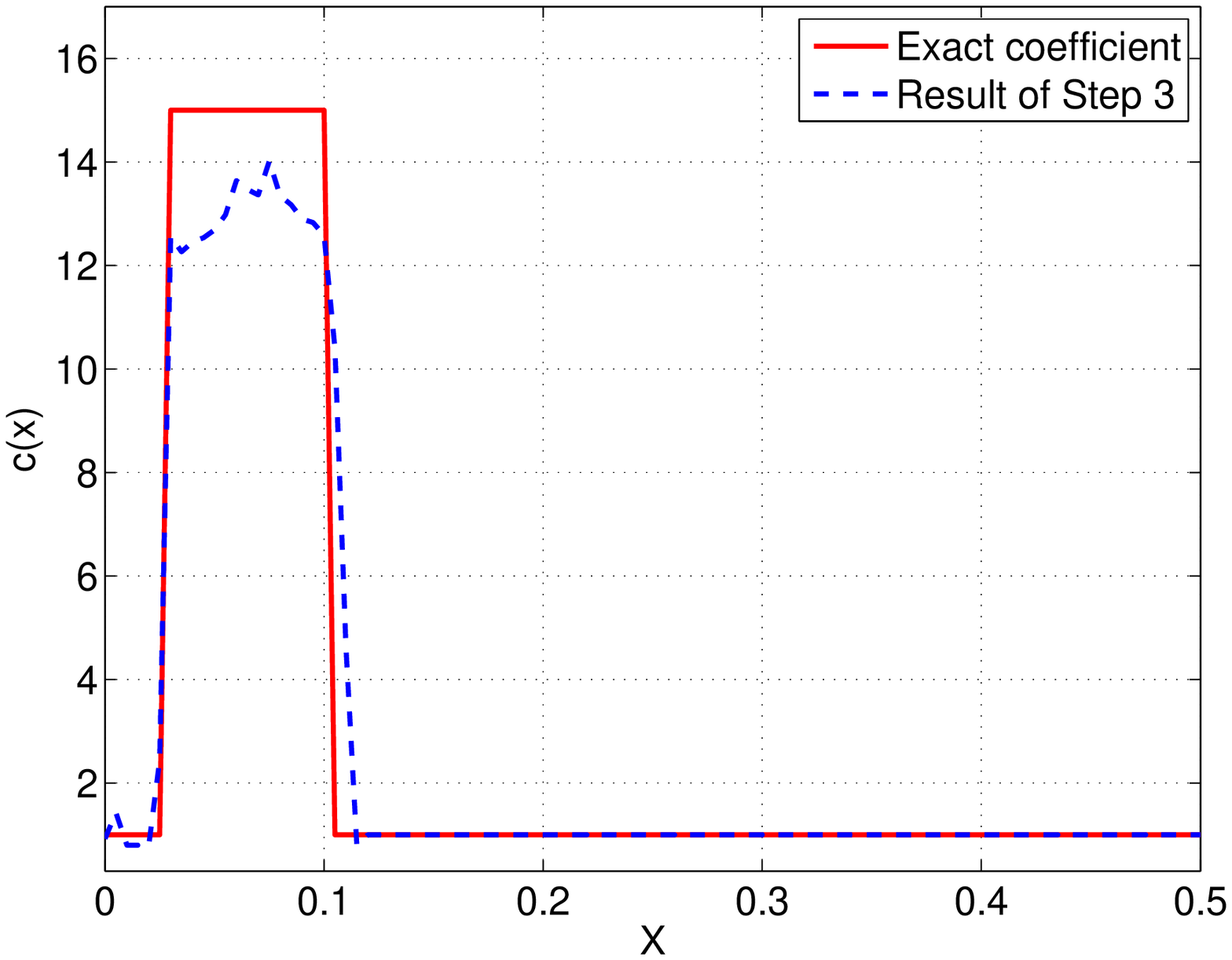}} 
\caption{Reconstruction of the coefficient of Example 2. (a) Result of Steps
1 and 2, (b) Result of Step 3. }
\label{fig:exa2}
\end{figure}

\noindent \textbf{Example 1.} Consider a piecewise constant coefficient
given by $c(x)=1+3\chi \lbrack 0.03,0.1],$ where $\chi $ is the
characteristic function. Figure \ref{fig:exa1} (a) compares the computed
coefficient values of Steps 1 and 2 with the exact one and Figure \ref%
{fig:exa1} (b) depicts the result of Step 3. To compare with the performance
of the above locally convergent method alone, we plot in Figure \ref%
{fig:exa1} (c), (d) the results of Steps 2 and 3, respectively, starting
from the homogeneous medium as the initial guess. We can see that our hybrid
algorithm provided accurate results, whereas the locally convergent method
alone failed.

\noindent \textbf{Example 2.} In this example, we consider another piecewise
constant coefficient with a larger jump $c(x)=1+14\chi \lbrack 0.03,0.1].$
The results of Steps 1 - 3 are shown in Figure \ref{fig:exa2}. Even though
the jump of the coefficient is high in this case, we still can see the good
accuracy of the reconstruction using our hybrid algorithm.

\begin{figure}[tph]
\centering
\subfloat[]{\includegraphics[width=0.42\textwidth,height =
0.32\textwidth]{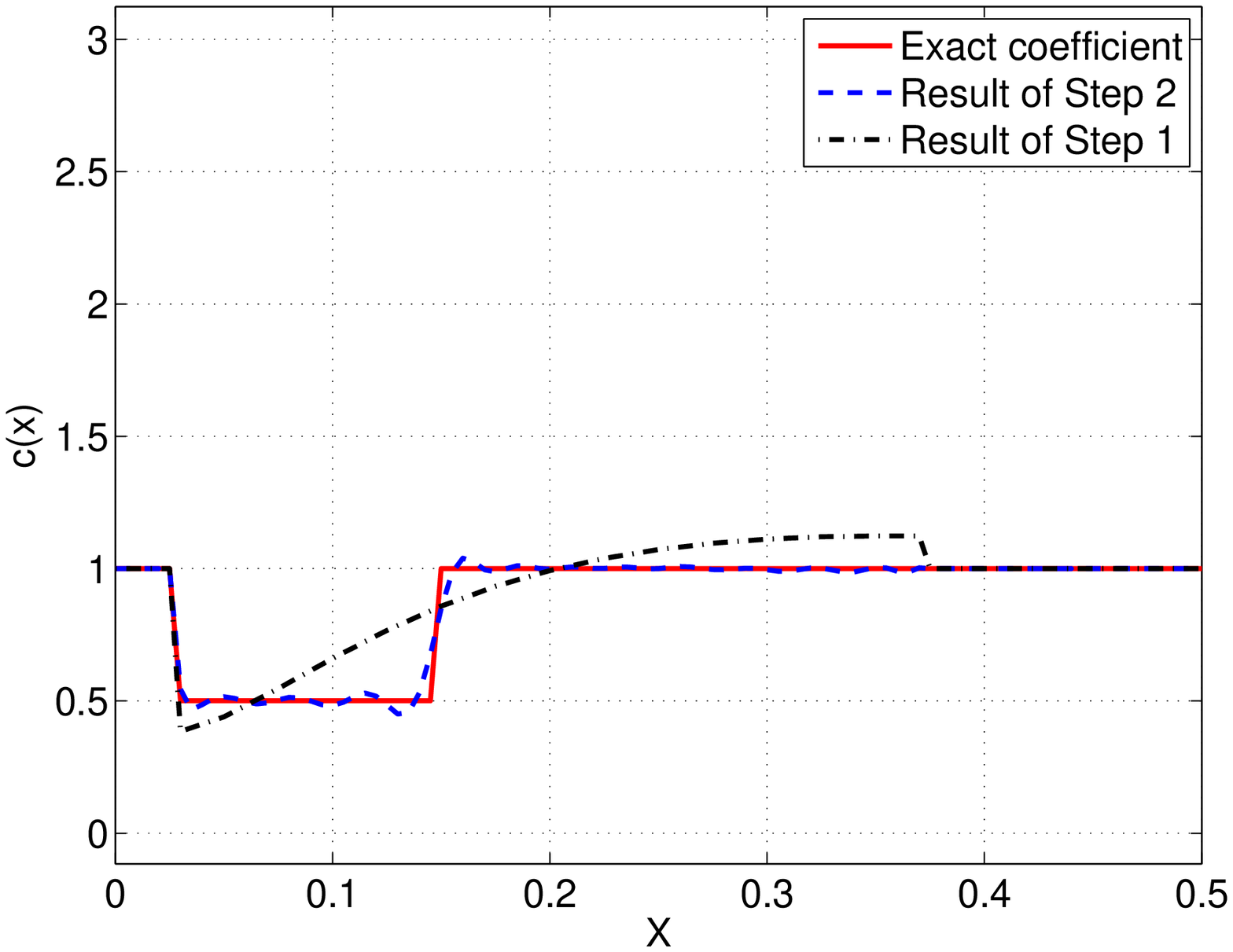}} 
\subfloat[]{\includegraphics[width=0.42\textwidth,height =
0.32\textwidth]{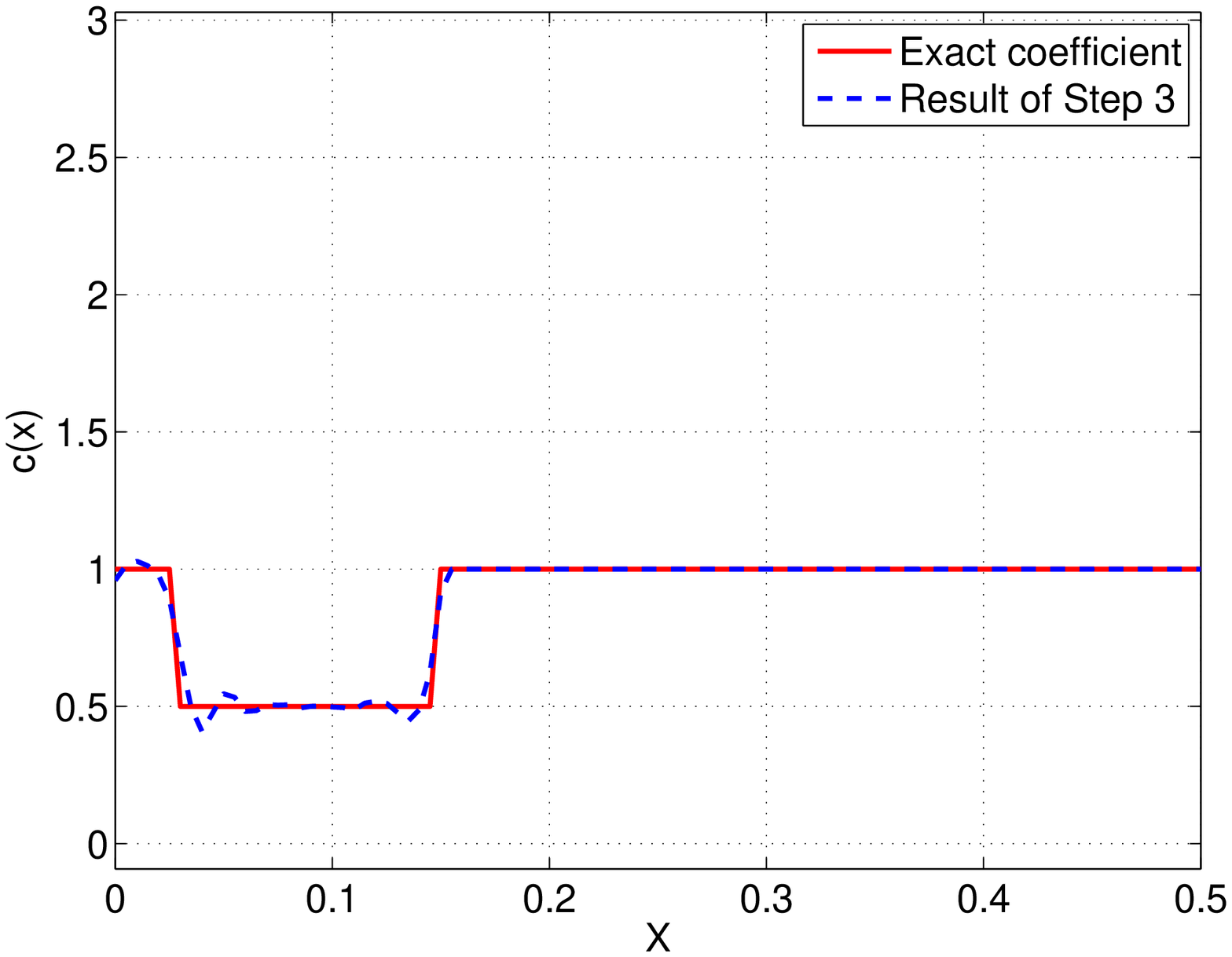}} 
\caption{Reconstruction of the coefficient of Example 3. (a) Result of Steps
1 and 2, (b) Result of Step 3. }
\label{fig:exa3}
\end{figure}

\begin{figure}[tph]
\centering
\subfloat[]{\includegraphics[width=0.42\textwidth,height =
0.32\textwidth]{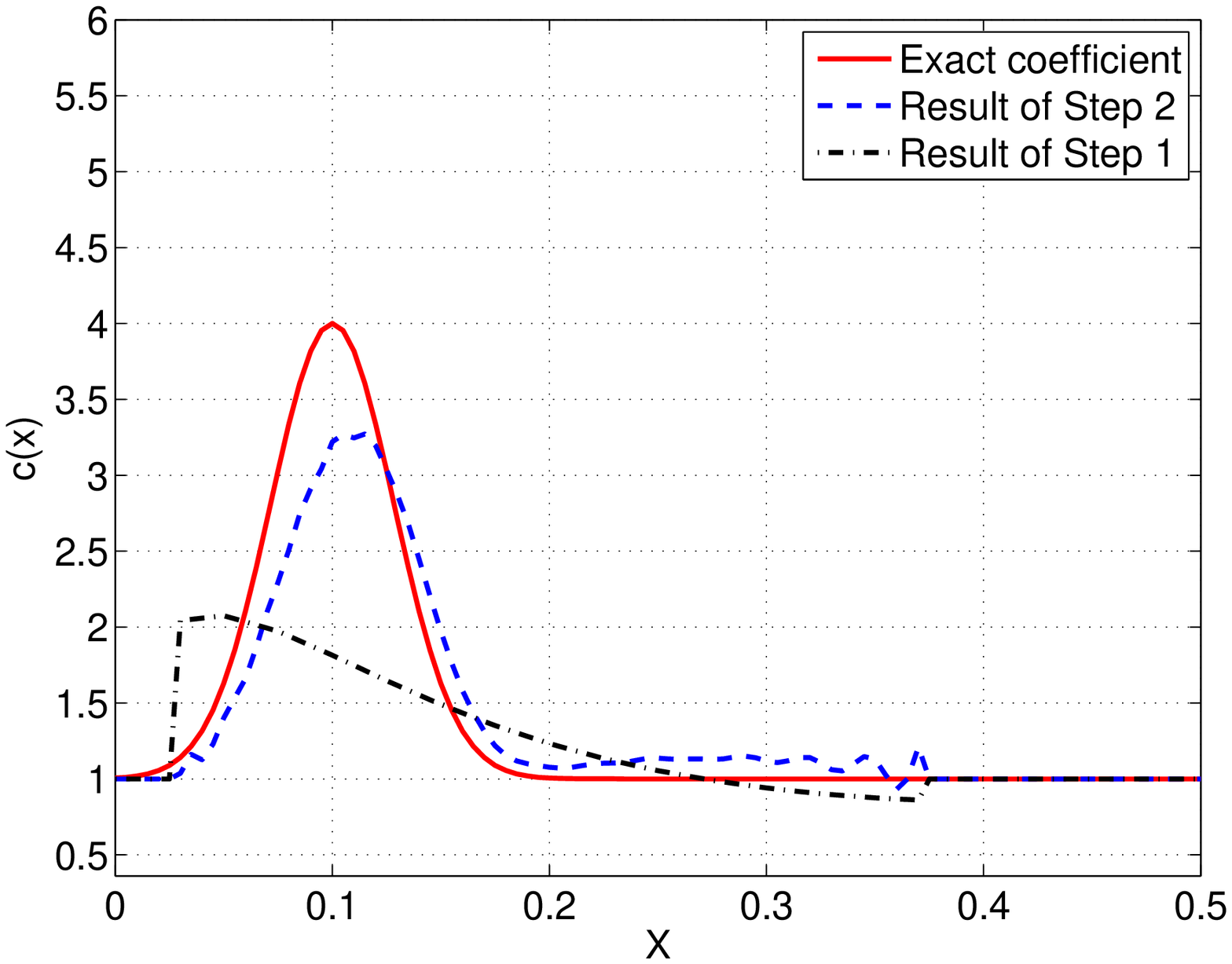}} 
\subfloat[]{\includegraphics[width=0.42\textwidth,height =
0.32\textwidth]{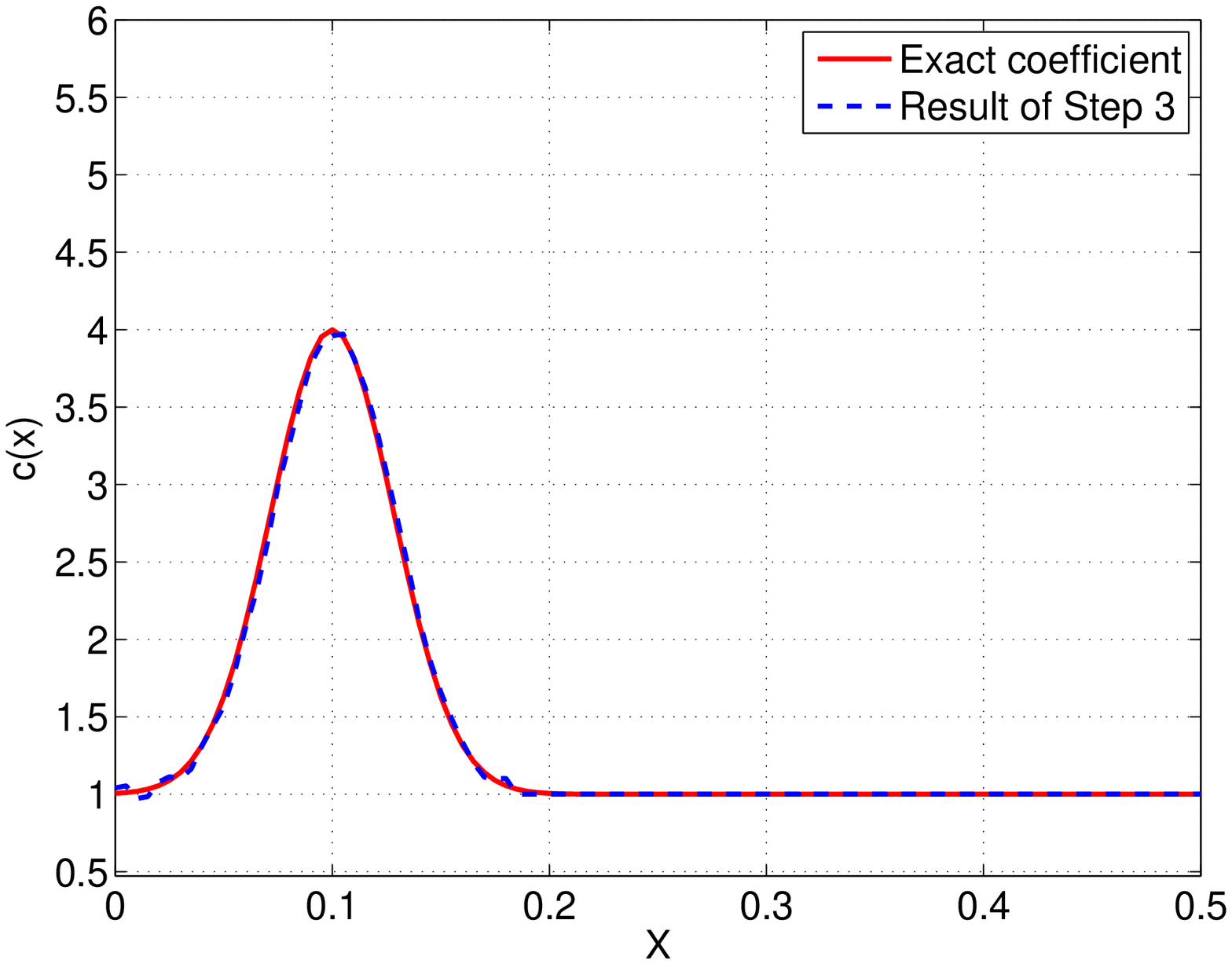}}
\par
\subfloat[]{\includegraphics[width=0.42\textwidth,height =
0.32\textwidth]{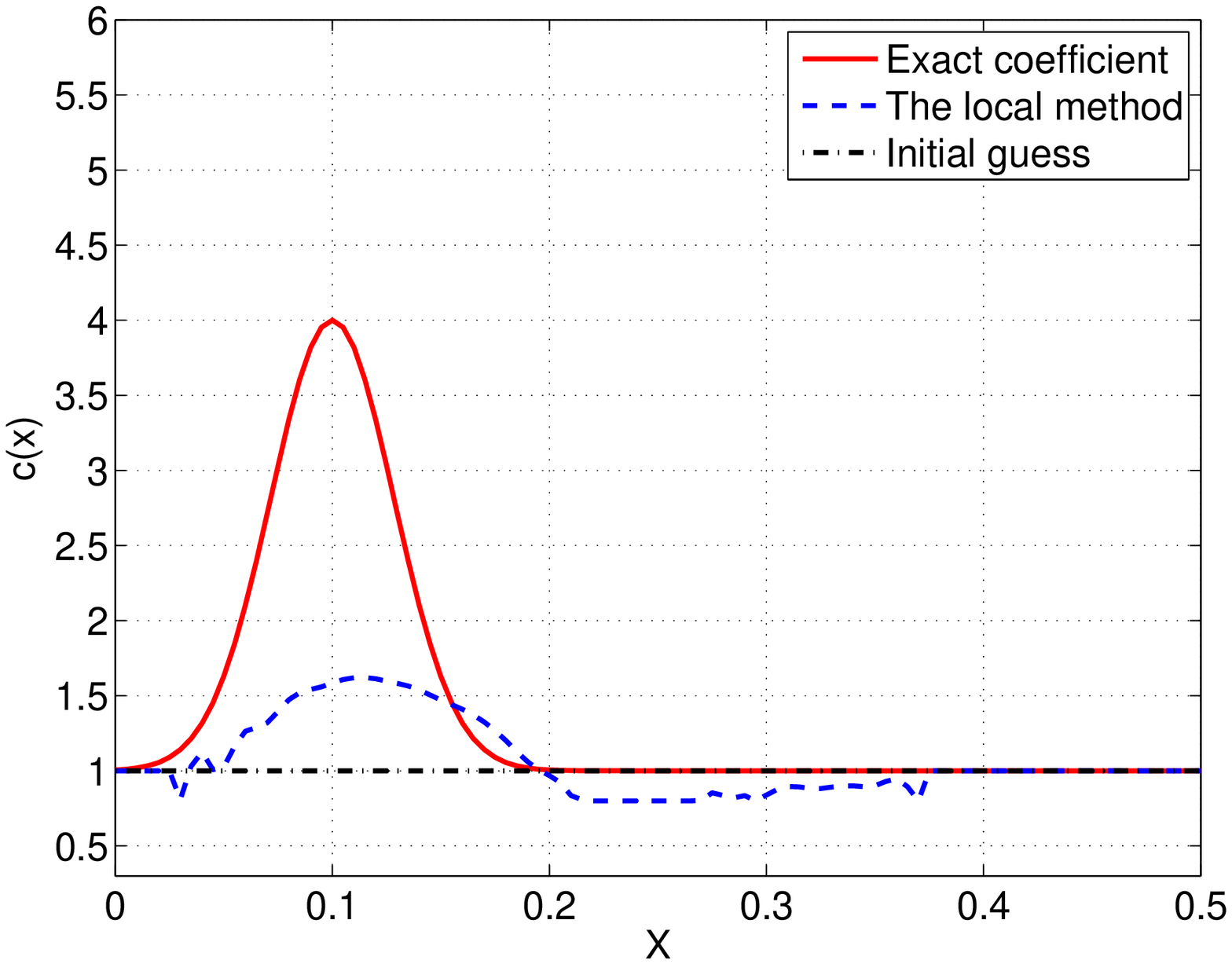}} 
\subfloat[]{\includegraphics[width=0.42\textwidth,height =
0.32\textwidth]{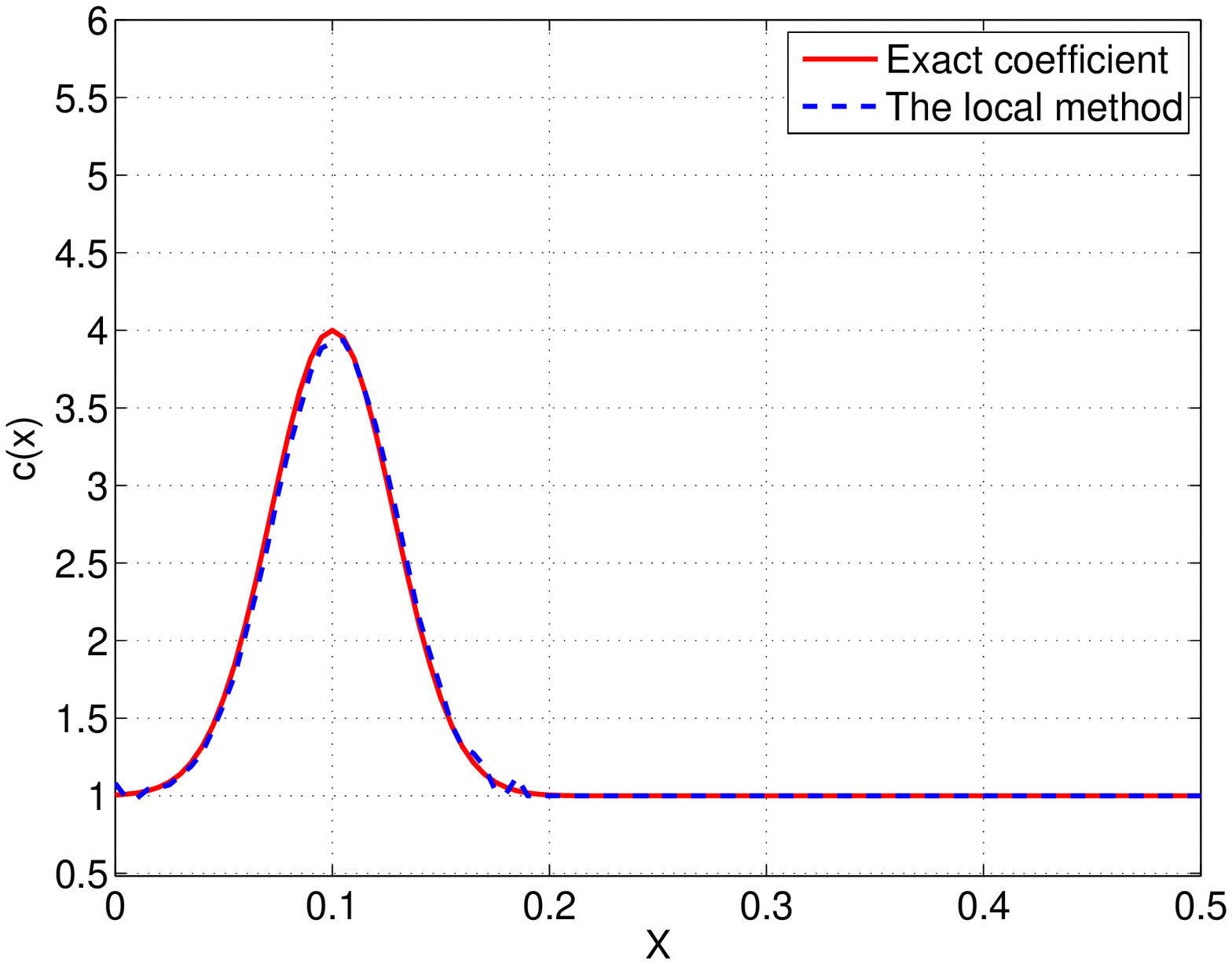}}
\caption{Reconstruction of the coefficient of Example 4. (a) Result of Steps
1 and 2, (b) Result of Step 3, (c) Result of Step 2 starting from the
homogeneous medium as the first guess, (d) Result of Step 3 applied to the
result of (c). }
\label{fig:exa4}
\end{figure}

\noindent \textbf{Example 3.} Consider the exact coefficient $c(x)=1-0.5\chi
\lbrack 0.03,0.15].$ This coefficient mimics the case the case when the
dielectric constant of an explosive is less than the one of a homogeneous
background. Figure \ref{fig:exa3} shows the reconstruction results of Steps
1 - 3 of the algorithm. Again, we obtained an accurate reconstruction.

\noindent \textbf{Example 4.} Finally, we consider a continuous coefficient
given by $c(x)=1+3e^{-(x-0.1)^{2}/\left( 0.04\right) ^{2}}.$ The results are
shown in Figure \ref{fig:exa4}. Comparing Figure \ref{fig:exa4} (a) with
Figure \ref{fig:exa4} (c), one can see that the combination of Step 1 and
Step 2 provided much better result than Step 2 starting from the homogeneous
medium as the first guess. However, results of Step 3 of both cases are
accurate.

From Figure \ref{fig:exa1} (d) and Figure \ref{fig:exa4} (d) we see that the
above locally convergent method, taking alone, is unstable in the sense
that, depending on the type of the target, it provides either bad or good
quality images. Meanwhile, the proposed hybrid algorithm provides accurate
results in all four examples.


\end{document}